\documentclass[12pt]{article}

\usepackage{amsfonts}
\usepackage{amsmath}
\usepackage{epsfig}
\usepackage{latexsym}
\usepackage{graphicx}
\usepackage{subfigure}
\usepackage{amssymb}
\numberwithin{equation}{section}
\allowdisplaybreaks

\def\spa#1{\phantom{\fbox{\rule[-#1cm]{0cm}{0cm}}}}

\def\be{\begin{equation}}
\def\ee{\end{equation}}
\def\bea{\begin{eqnarray}}
\def\eea{\end{eqnarray}}
\def\bequ{\begin{equation}}
\def\eequ{\end{equation}}

\def\half{{1\over 2}}

\def\del{\partial}
\def\nn{\nonumber}

\renewcommand{\thefootnote}{\fnsymbol{footnote}}

\begin{document}

\hfuzz=100pt
\title{{\Large \bf{On the scalar graviton in $n$-DBI gravity}}}
\date{}
\author{Fl\'avio S. Coelho$^a$\footnote{flavio@physics.org}, Carlos Herdeiro$^a$\footnote{herdeiro@ua.pt}, Shinji Hirano$^{b}$\footnote{hirano@eken.phys.nagoya-u.ac.jp} and Yuki Sato$^{b,c}$\footnote{ysato@th.phys.nagoya-u.ac.jp}
  \spa{0.5} \\
\\
$^a${\small{\it Departamento de F\'isica da Universidade de Aveiro and I3N}}
\\ {\small{\it Campus de Santiago, 3810-183 Aveiro, Portugal}}\\
$^{b}${\small{\it Department of Physics, Nagoya University}}
\\ {\small{\it Nagoya 464-8602, Japan}}\\
$^{c}${\small{\it The Niels Bohr Institute}}
\\ {\small{\it Blegdamsvej 17, DK-2100, Copenhagen, Denmark}}
}
\date{\small{May 2012}}

\maketitle
\centerline{}

\begin{abstract}
$n$-DBI gravity is a gravitational theory which yields near de Sitter inflation spontaneously at the cost of breaking Lorentz invariance by a preferred choice of foliation. We show that this breakdown endows $n$-DBI gravity with one extra physical gravitational degree of freedom: a \textit{scalar graviton}. Its existence is established by Dirac's theory of constrained systems. Firstly, studying scalar perturbations around Minkowski space-time, we show that there exists one scalar degree of freedom and identify it in terms of the metric perturbations. Then, a general analysis is made in the canonical formalism, using ADM variables. It is useful to introduce an auxiliary scalar field, which allows recasting $n$-DBI gravity in an Einstein-Hilbert form but in a Jordan frame. Identifying the constraints and their classes we confirm the existence of an extra degree of freedom in the full theory, besides the two usual tensorial modes of the graviton. We then argue that, unlike  the case of (the original proposal for) Ho\v{r}ava-Lifschitz gravity, there is no evidence that the extra degree of freedom originates pathologies, such as vanishing lapse, instabilities and strong self-coupling at low energy scales. 

\end{abstract}

\renewcommand{\thefootnote}{\arabic{footnote}}
\setcounter{footnote}{0}

\newpage
\tableofcontents

\section{Introduction}
General diffeomorphism invariance is a central property of General Relativity (GR). In particular, it means that all foliations of space-time (say by space-like hypersurfaces) are equivalent. For many pratical purposes a choice of foliation must be made, which can be achieved, for instance, by an ADM decomposition of space-time \cite{Arnowitt:1960es,York:1972sj}. The ten metric degrees of freedom are then decomposed into the lapse, shift and 3-metric, of which only the latter contains physical information. The lapse and the shift are pure gauge, as it is manifest in the fact that their equations of motion are actually constraints. Such constraints need only be imposed in one leaf of the foliation and are guaranteed to be preserved in the evolution. The number of physical degrees of freedom of the theory can be formally computed by Dirac's theory of constrained systems \cite{Dirac:1950pj}. For GR there are 20 phase space variables and 8 first class constraints, leading to $(20-2\times 8)/2$ degrees of freedom. These are the two polarisations of the graviton, which are clearly seen by studying plane waves in linearised theory,  corresponding to trace-free, transverse, tensor modes.

Any theory of gravity for which general diffeomorphism invariance is broken will have, generically, a different number of physical degrees of freedom. An example is when the full diffeomorphism invariance is broken to foliation preserving diffeormorphisms (FPD). This type of symmetry has been extensively discussed in recent years in the context of Ho\v{r}ava-Lifschitz (HL) gravity \cite{Horava:2009uw}. Under this smaller symmetry, the shift can still be gauged away, but the lapse (and the conformal mode, cf. Section \ref{312}) acquires physical significance. As a manifestation, the equation of motion of the lapse, the Hamiltonian constraint  $\mathcal{H}$, is not preserved in the evolution: the evolution of $\mathcal{H}$ is actually a dynamical equation for the lapse. Naively, this seems to introduce one new degree of freedom. In the non-projectable class of HL theories, however, a formal counting, based on the fact that the Hamiltonian constraint becomes second class and the lapse is determined (removing two phase space variables), gives $(2\times (6+3)-2\times (3+3)-1)/2=5/2$ degrees of freedom \cite{Henneaux:2009zb}. Getting one half degree of freedom might not necessarily be problematic in itself.\footnote{A well-known example is a chiral boson.} There are, however, at least three problematic features related to the existence of this new half degree of freedom in HL gravity.

The first problem concerns the absence of dynamics and is associated to the fact that the extra degree of freedom is halved. For generic asymptotically flat space-times, the lapse is forced to be zero at spatial infinity \cite{Henneaux:2009zb}. Actually, for particular values of the couplings, it was shown (and suggested this could be the case for generic values of the couplings) that the lapse must vanish everywhere. This indicates that there is no dynamics in HL gravity.
The second problem is the short distance instability the scalar mode might trigger (if the dynamics was not frozen). Looking at perturbations around generic backgrounds, it was found in \cite{Blas:2009yd} that the high frequency modes of the extra degree of freedom develop an imaginary part and the perturbations can grow very swiftly in time.\footnote{The first two issues are apparently contradictory; if the lapse must collapse everywhere, there would be no way to develop an exponentially growing mode involving the lapse. We will, however, discuss the instabilities of the model which evades the first issue. }
The third problem is the self-coupling of the scalar mode which remains strong to very low energy scales \cite{Blas:2009yd, Charmousis:2009tc}. This implies that the extra scalar mode never decouples and thus HL gravity does not flow to GR in the IR as it was hoped.

Recently, another model of gravity for which the full diffeomorphism invariance is broken to FPD was proposed by some of us \cite{Herdeiro:2011km,Herdeiro:2011im}: $n$-DBI gravity. This model was designed to reduce to the Dirac-Born-Infeld (DBI) scalar field theory for conformally flat geometries; it was motivated by the observation that for such geometries, in a cosmological setup, two epochs of acceleration for the universe are naturally obtained  \cite{Herdeiro:2011km}. Subsequently, we observed in  \cite{Herdeiro:2011im} that a large class of exact solutions of GR (including the most standard spherical black hole solutions, in the most standard foliations) are solutions of $n$-DBI gravity.

The purpose of the present paper is twofold. Firstly, to understand what are the extra degrees of freedom present in $n$-DBI gravity due to the breakdown of full diffeomorphism invariance. Secondly, to investigate whether $n$-DBI gravity is afflicted by problems similar to those described above in the context of HL gravity.

Concerning the first part of our investigation, we shall show that $n$-DBI gravity possesses one extra degree of freedom, as compared to GR, that this degree of freedom is a scalar graviton and that its existence is a direct consequence of the breakdown of the full diffeomorphisms group to  FPD. This shall be done in Sections 3 and 4, first by considering perturbations around Minkowski spacetime and then by considering the fully non-linear $n$-DBI theory. Concerning the second part of our investigation, we shall present arguments, in Section 5, that none of the aforementioned problems that afflict HL gravity are present in $n$-DBI. This healthy behaviour is intimately connected with the nonlinear lapse dependence in $n$-DBI, as compared to HL gravity. Indeed, $n$-DBI gravity naturally contains lapse dependent terms in the action, analogous to those suggested in \cite{Blas:2009qj} for the consistent extension of HL gravity.  

The analysis in Sections 4 and 5 can be made simpler by introducing alternative forms for $n$-DBI gravity. One alternative form linearizes the action, i.e. eliminates the awkward square root dependence, by introducing an auxiliary field. Another alternative form makes the action covariant, by introducing a St\"uckelberg field. These alternative forms are introduced in Section 2. We shall conclude, in Section 6, by discussing more on the stability and commenting on future research directions for understanding this model. Some technicalities are presented in three Appendices.

\section{Different forms of $n$-DBI gravity}
The action for $n$-DBI gravity without matter is \cite{Herdeiro:2011km,Herdeiro:2011im}
\begin{align}
S=-{3\lambda\over 4\pi G_N^2}\int d^4x\sqrt{-g}\left\{\sqrt{1+{G_N\over 6\lambda}\left(\mbox{}^{\mbox{}^{(4)}\!}\! R+\mathcal{K}\right)}-q\right\}\ ,
\label{action}
\end{align}
where $G_N$ is Newton's constant, $q,\lambda$ are the two dimensionless parameters of the theory and $\mbox{}^{\mbox{}^{(4)}\!}\! R$ is the four dimensional Ricci scalar. To completely define the theory a foliation structure must be chosen. Let ${\bf n}$ be a unit time-like vector field, everywhere orthogonal to the leaves of such foliation; let $h_{\mu\nu}=g_{\mu\nu}+n_\mu n_\nu$ be the first fundamental form and $K_{\mu\nu}=\frac{1}{2}\mathsterling_{\bf n} h_{\mu\nu}$ the second fundamental form. Then $\mathcal{K}$ in (\ref{action}) is defined as
\bequ
\mathcal{K}\equiv -\frac{2}{\sqrt{h}}\mathsterling_{\bf n}(\sqrt{h}K) \ , \qquad K\equiv K_{\mu\nu}h^{\mu\nu} \ .
\eequ

Performing the Arnowitt-Deser-Misner (ADM) decomposition \cite{Arnowitt:1960es}
\be
ds^2=-N^2 dt^2+h_{ij}\left(dx^i+N^i dt\right)\left(dx^j+N^j dt\right)\ ,
\ee
where $N$ and $N_i$, respectively, are the lapse and shift functions, and $h_{ij}$ with $i, j=1, 2, 3$ is the spatial metric, the action (\ref{action}) becomes
\bequ
S=-{3\lambda\over 4\pi G_N^2}\int dtd^3x\sqrt{h}N\biggl[\sqrt{1+{G_N\over 6\lambda}\left(R+K_{ij}K^{ij}-K^2-2N^{-1}\Delta N\right)}-q\biggr]\label{DBIgravity} \ , 
\eequ
where $R$ is the Ricci scalar of $h_{ij}$. In particular, observe the lapse term $N^{-1}\Delta N$. As suggested in \cite{Blas:2009qj}, adding analogous terms can provide a consistent extension of HL gravity. Indeed, as we shall see below, this term plays a crucial role in evading the pathologies that afflicted HL gravity.

\subsection{A linearised form with auxiliary field}
\label{auxfield}
As it is known in various contexts, square root type actions may be linearised by introducing auxiliary fields. A well known example is the classical equivalence between the Polyakov and Nambu-Goto actions in string theory, via the introduction of an auxiliary metric (the world sheet metric). A similar reformulation for the Eddington inspired Born-Infeld gravity was provided recently \cite{Delsate:2012ky}, for which the auxiliary variable is the ``apparent" metric. In our case, with the introduction of an auxiliary scalar field $e$, the action (\ref{action}) becomes\footnote{We thank Soo-Jong Rey and Takao Suyama for suggesting this formulation.}
\begin{equation}
S^{e}=-\frac{1}{16\pi G_N}\int d^4x\sqrt{-g}e\left[\mbox{}^{\mbox{}^{(4)}\!}\! R-2G_N\Lambda_C(e)+\mathcal{K}\right],\qquad \Lambda_C(e)=\frac{3\lambda}{G_N^2}\left(\frac{2q}{e}-1-\frac{1}{e^2}\right) \ .
\label{linearisedaction}
\end{equation}
For constant field $e$, this is the Einstein-Hilbert action with a cosmological constant (in exactly the form found in \cite{Herdeiro:2011im} in terms of the integration constant $C$ therein) with the Gibbons-Hawking-York boundary term \cite{Gibbons:1976ue,York:1972sj}. For generic $e$, the theory resembles General Relativity in a Jordan frame (but indeed it is quite different). This form will be used in Section \ref{canfor} for the canonical formalism analysis of $n$-DBI gravity.

It is worth noting that we can rewrite the action (\ref{linearisedaction}) in the Einstein frame by performing a Weyl transformation: 
\bequ
g_{\mu\nu}\rightarrow e^{-1}g_{\mu\nu} \ .
\eequ
Redefining the auxiliary field $e\equiv \exp{(2\chi)}$, the action becomes, up to boundary terms,
\begin{align}
S_{\rm Einstein}=-\frac{1}{16\pi G_N}\int d^4x\sqrt{-g}\left[\mbox{}^{\mbox{}^{(4)}\!}\! R-6\left(n^\alpha n^\beta+h^{\alpha\beta}\right)\del_\alpha\chi\del_\beta\chi+2{\cal K}\chi+V(\chi)\right] \ ,
\label{Einstein}
\end{align}
with the potential
\begin{align}
V(\chi)=\frac{6\lambda}{G_N}\exp{(-4\chi)}\left[\left(\exp{(\chi)}-\exp{(-\chi)}\right)^2+2(1-q)\right]\ .\label{potential}
\end{align}
There is a caveat: it is misleading to regard the Einstein frame theory as a scalar-tensor theory (the  scalar being $\chi$). Despite the appearance of the kinetic term, the scalar field $\chi$ is still an auxiliary field and does not give rise to an {\it independent} degree of freedom. 
As we will see, the extra scalar mode is furnished in the metric and the scalar field $\chi$ is only related to it through the equations of motion.

\subsection{A covariant form with St\"uckelberg field}
$n$-DBI gravity breaks Lorentz invariance due to the coupling of gravity to the unit time-like vector field ${\bf n}$, which defines a preferred space-time foliation. Full general covariance can be restored by introducing a St\"uckelberg field $\phi(x^\mu)$, such that its gradient is everywhere time-like and non-vanishing:
\bequ
n_\mu=-\frac{\partial_\mu\phi}{\sqrt{-X}} \ , \qquad g^{\mu\nu}n_{\mu}n_{\nu}=-1\ , \qquad X\equiv g^{\mu\nu}\partial_\mu\phi\partial_\nu\phi \ .\label{def_phi}
\eequ
By definition the theory is invariant under $\phi\to f(\phi)$, where $f(\phi)$ is an arbitrary function of $\phi$. Note that the original non-covariant form is recovered when $\phi=t$ for which ${\bf n}=(-N, 0, 0, 0)$, and the symmetry reduces to the time reparametrization $t\to f(t)$.
A similar treatment in the case of HL gravity has been performed in \cite{Blas:2009yd}. The extrinsic curvature becomes
\bequ
K_{\mu\nu}
=\half\left[n^{\alpha}D_{\alpha}h_{\mu\nu}+h_{\mu\alpha}D_{\nu}n^{\alpha}
+h_{\nu\alpha}D_{\mu}n^{\alpha}\right]\ ,\qquad\quad
K=D_{\alpha}n^{\alpha}\ .
\eequ
It then follows that
\bequ
{\cal K}=-2D_{\alpha}\left(n^{\alpha}D_{\beta}n^{\beta}\right) \ .
\eequ
Thus, the $n$-DBI action may be rewritten in the covariant form
\begin{align}
S_{\rm c}=-{3\lambda\over 4\pi G_N^2}\int d^4x\sqrt{-g}\left\{\sqrt{1+{G_N\over 6\lambda}\left(\!\!\mbox{}^{\mbox{}^{(4)}\!}\! R-2D_{\alpha}\left[\frac{\partial^\alpha\phi}{\sqrt{-X}} D_{\beta}\left(\frac{\partial^\beta\phi}{\sqrt{-X}} \right)\right]\right)}-q\right\}\ .\label{DBIgravitycovariant}
\end{align}
This form will be used in the discussions of Section \ref{pathologies}. 
In contrast to the Einstein frame theory, this covariant theory can be thought of as a scalar-tensor theory. Indeed, in this form, the scalar field $\phi$ does yield an independent degree of freedom. Put differently, the scalar mode in the metric is entirely transferred to the St\"uckelberg field $\phi$.

Clearly, this can be linearised again by introducing the auxiliary field $e$,
\begin{align}
S_{\rm c}^e=-\frac{1}{16\pi G_N}\int d^4x\sqrt{-g}e\left\{\!\mbox{}^{\mbox{}^{(4)}\!}\! R-2G_N\Lambda_C(e)
-2D_{\alpha}\left[\frac{\partial^\alpha\phi}{\sqrt{-X}} D_{\beta}\left(\frac{\partial^\beta\phi}{\sqrt{-X}} \right)\right]\right\}\ .\label{DBIgravitycovariante}
\end{align}
In these formulations, general covariance gets {\it spontaneously} broken by the scalar field $\phi$ acquiring the vev $\langle\phi\rangle=t$. It should be noted that this has a certain bearing on the ghost condensation of \cite{ArkaniHamed:2003uy}.

\section{The scalar graviton}
In this section we shall show that there are three physical degrees of freedom in $n$-DBI gravity: the two usual tensorial modes of general relativity plus one \lq\lq scalar graviton". This is in contrast with HL gravity where there are two and a half physical degrees of freedom \cite{Henneaux:2009zb}.

\subsection{Scalar perturbations around flat space-time}
We shall first study the scalar mode in flat space-time. In particular, we shall clarify how the scalar mode becomes physical as a direct consequence of the breakdown of full diffeomorphism invariance to FPD. 

\subsubsection{The existence of a scalar graviton}
\label{gravexists}

The metric has ten components; the lapse $N$, the shift $N^i$, and the spatial metric $h_{ij}$. These can be decomposed into 4 scalars $(n, B, \psi, E)$, 2 transverse vectors $(A_i, \tilde{A}_i)$, 1 transverse traceless tensor $\tilde{h}_{ij}$ as follows (see e.g. \cite{Stewart:1990fm}):  
\begin{align}
N=&1+n\ , \qquad\qquad N_i=\nabla_i B+A_i\ ,\nn\\
h_{ij}=&\delta_{ij}-2\left(\delta_{ij}-{\nabla_i\nabla_j\over\Delta}\right)\psi
-2{\nabla_i\nabla_j\over\Delta} E
+\left(\nabla_i\tilde{A}_j+\nabla_j\tilde{A}_i\right)+\tilde{h}_{ij}\ ,
\label{metricdecomp}
\end{align}
where the transversality and traceless conditions are imposed,
\begin{align}
\nabla^iA_i=\nabla^i\tilde{A}_i=\nabla^i\tilde{h}_{ij}=\tilde{h}^i_{\mbox{ }i}=0\ .
\label{transverse}
\end{align}

We now expand the $n$-DBI gravity action (\ref{DBIgravity}) to quadratic order around flat space-time, setting $q=1$. Thanks to conditions (\ref{transverse}), scalar perturbations and vector-tensor perturbations decouple from each other. The quadratic Lagrangian density for the scalar fields yields, up to total divergences, 
\begin{align}
4\pi G_N{\cal L}_{\rm scalar}=2\dot\psi^2+4\dot{\psi}\left(\dot{E}+\Delta B\right)+\left(2\psi-4n\right)\Delta\psi
+{G_N\over6\lambda}\left(2\Delta\psi-\Delta n\right)^2\ .
\label{scalarL}
\end{align}
This essentially corresponds to the $\lambda=1$ case of \cite{Blas:2009qj}, and the scalar graviton in this model is qualitatively different in its character from the one discussed there.

In order to establish the existence of the scalar degree of freedom, 
we shall apply Dirac's theory of constrained systems. The Hamiltonian density of this effective scalar theory reads\footnote{For convenience, we have  rescaled the scalar fields by the factor of $(4\pi G_N)^{1/2}$.}
\begin{align}
{\cal H}^{(0)}_{\rm scalar}=-{1\over 8}p_E^2+{1\over 4}p_Ep_{\psi}-p_E\Delta B
-\left(2\psi-4n\right)\Delta\psi
-{G_N\over6\lambda}\left(2\Delta\psi-\Delta n\right)^2\ ,
\end{align}
where $p_E$ and $p_{\psi}$ are the conjugate momenta of $E$ and $\psi$, respectively. 
Since the Lagrangian density (\ref{scalarL}) does not contain the time derivative of $n$ or $B$, their conjugate momenta become the following primary constraints:
\begin{align}
\Phi_1\equiv p_{n}=0\ ,\qquad
\Phi_2\equiv p_B=0\ .
\end{align}
Thus the dynamics of this system is governed by the Hamiltonian density
\begin{align}
{\cal H}^{(1)}_{\rm scalar}={\cal H}^{(0)}_{\rm scalar}+\lambda_1\Phi_1+\lambda_2\Phi_2\ ,
\label{hamper3}
\end{align}
where $\lambda_1$ and $\lambda_2$ are Lagrange multipliers.
The consistency requires the time flows of the primary constraints to vanish  (herein $\approx$ means `weakly equal' as standard in Dirac's theory):
\begin{align}
\Phi_4\equiv \dot{\Phi}_1= 4\Delta\psi+{G_N\over 6\lambda}\left(4\Delta^2\psi -2\Delta^2n\right)
\approx 0\ ,\qquad
\Phi_5\equiv \dot{\Phi}_2=-\Delta p_E\approx 0\ .
\end{align}
The former corresponds to the Hamiltonian constraint and the latter to the momentum constraint, as will become clear in the next section.  
The time flows of these secondary constraints do not yield new constraints.

Among these four constraints, $\Phi_2$ and $\Phi_5$ commute with all the constraints and are therefore first class, whereas $\Phi_1$ and $\Phi_4$ are second class. Hence the physical degrees of freedom of this system are counted as $(2\times 4 - 2 \times 2 - 2)/2 = 1$. In contrast, in GR where $\lambda\to \infty$, all the constraints are first class and the counting becomes $(2\times 4 - 2 \times 2 - 2\times 2)/2 = 0$. In other words, GR has two pairs of first class constraints associated with the full diffeomorphism, whereas $n$-DBI has only one pair reflecting that only FPD is preserved. Thus, in $n$-DBI gravity there remains one physical scalar degree of freedom that cannot be gauged away. We will elaborate on this point in a moment.

Note that it is the presence of the lapse term $(\Delta n)^2\sim (N^{-1}\Delta N)^2$ in the Hamiltonian that leaves one physical scalar degree of freedom, as opposed to $1/2$ in HL gravity.\footnote{In non-projectable HL gravity, the lapse $N$ is a Lagrange multiplier as in GR, but the Hamiltonian constraint is second class. Moreover, the time flow of the Hamiltonian constraint yields an additional constraint that depends on the lapse $N$. Thus, there are 3 second class constraints; the conjugate momentum $p_N$ of the lapse, the Hamiltonian constraint ${\cal H}$, and its time flow $\dot{\cal H}$. Together with 6 first class constraints, the number of degrees of freedom is counted as $(2\times 10-2\times 6-3)/2=2+1/2$. 
Note, however, that linear perturbations about flat space-time yield a misleading result. There appear four second and two first class constraints, implying incorrectly that the number of scalar degrees of freedom is zero.
In projectable HL gravity, there is an additional primary constraint $\del_iN=0$ on top of $p_N=0$. In this case, the time flow of $p_N=0$ does not yield the Hamiltonian constraint. Instead, it determines the Lagrange multiplier of $\del_i N$. Hence $p_N$ and $\del_iN$ are the only second class constraints, and the total physical degrees of freedom is $2+1$.} This is similar to the consistent extension of HL gravity \cite{Blas:2009qj}.

\subsubsection{The identity of scalar graviton}
\label{312}

Having established the existence of the scalar graviton, we wish to identify it more explicitly in terms of the metric perturbations.  For this purpose, we first note that the scalar field theory (\ref{scalarL}) has the FPD gauge symmetry:
\begin{align}
\psi&\to\psi\ , \label{FPDlinear1}\\
B&\to B+\dot{L}-T\ ,\\
E&\to E-\Delta L\ ,\\
n&\to n+\dot{T}\ ,
\label{FPDlinear4}
\end{align}
where $T$ and $L$ are related to infinitesimal coordinate transformations  by $\xi^0=T$ and $\xi^i=\nabla^iL$. In $n$-DBI gravity, $T$ is a function of time only, whereas $L$ is a function of both space and time. In GR, $T$ also becomes a function of both space and time.
Note that the gauge invariant quantities are
\begin{align}
\{\psi, \ddot{E}+\Delta \dot{B}+\Delta n\}\qquad\qquad &\mbox{for both GR and $n$-DBI}\ ,\\
\{\dot{E}+\Delta B,\del_i n \} \qquad\qquad\qquad &\mbox{for $n$-DBI only}\ .
\end{align}
As it is clear from the above counting of degrees of freedom, the extra scalar graviton exists in $n$-DBI gravity, because the constraints $\Phi_1$ and $\Phi_4$ are second class, whereas in GR they are first class and generate the gauge transformations
\bequ
\left\{ \int d^3y \zeta_1(y)\Phi_1(y),n(x)\right\}=\zeta_1(x) \ , \qquad \left\{ \int d^3y \zeta_2(y)\Phi_4(y),p_\psi(x)\right\}=-4\Delta \zeta_2(x) \ .
\eequ
Comparing this with the FPD (\ref{FPDlinear1})-(\ref{FPDlinear4}), we find that
\bequ
\zeta_1=\dot{T} \ , \qquad \zeta_2=T \ ,
\eequ
where we used $p_\psi=4(\dot{\psi}+\dot{E}+\Delta B)$. In $n$-DBI gravity, as we stressed above, $T$ is a function of time only. Accordingly, the constraints $\Phi_1$ and $\Phi_4$ become second class and are not considered as generators of gauge transformations. This implies that the scalar graviton should involve the lapse and/or the shift whose gauge transformations are generated by $T$. This nicely fits the expectation that the scalar graviton must have something to do with the foliation structure which is specified by the lapse and the shift. Indeed, the equations of motion, 
\begin{align}
\ddot{\psi}&=0\ ,\label{eom1}\\
\Delta\dot{\psi}&=0\ ,\label{eom2}\\
\ddot{E}+\Delta\dot{B}+\Delta n+\Delta\psi&=0\ ,\label{eom3}\\
\Delta\psi&={G_N\over 6\lambda}\left({1\over 2}\Delta^2n-\Delta^2\psi\right)\ ,\label{eom4}
\end{align}
have in the $E=0$ gauge the general solution\footnote{A more common gauge is to set the shift $B=0$. We can go from the $E=0$ to the $B=0$ gauge by choosing the gauge parameter $L(t,x)=-B_0(x) t-\half B_1(x) t^2$. This yields the conformal mode $E(t,x)=\Delta B_0(x) t +\half \Delta B_1(x) t^2$. Note that in either gauge the lapse $n$ alone only accounts for a half degree of freedom of the scalar graviton.}
\begin{align}
B(t, x)=&B_0(x) + B_1(x) t\ ,\label{sol1}\\
n(t,x)=&-B_1(x)-\psi_0(x)\ ,\label{sol2}\\
\psi(t,x)=&\psi_0(x)\ ,\label{sol3}
\end{align}
with $\psi_0(x)$ related to $B_1(x)$ by
\be
\psi_0(x)=-{G_N\over 6\lambda}\left(\Delta B_1(x)+{3\over 2}\Delta\psi_0(x)\right)\ .
\ee
Note that we have imposed the boundary condition that all the fields must fall off at spatial infinity. In other words, the functions $f(x)$'s appearing in the most general solution and obeying the Laplace equation $\Delta f(x)=0$ are unphysical and set to zero.

In the Hamiltonian system, a degree of freedom is the freedom to choose a pair of initial data for the time evolution in the phase space. We have found exactly one degree of freedom, {\it i.e.}, the initial data specified by a pair of arbitrary functions of space, $\left(B_0(x), B_1(x)\right)$.
In GR, these could have been gauged away by choosing the gauge parameter $T(t,x)=B(t,x)$ (and the Hamiltonian constraint would have enforced $\psi_0(x)=0$). Put differently, in $n$-DBI gravity the scalar mode is the broken gauge degree of freedom $T(t,x)$ which obeys, by taking the time derivative of the the Hamiltonian constraint (\ref{eom4}),
\be
\Delta^2\ddot{T}(t,x)=0\ ,
\label{scalargravitoneq}
\ee
where we have used $n(t,x) = - \dot{T}(t,x) -\psi_0(x)$ and $\psi(t,x)=\psi_0(x)$. 
A few remarks are in order: firstly, this is a key equation, despite its extremely simple appearance, and notably originates from the nonlinear lapse term $(\Delta n)^2\sim (N^{-1}\Delta N)^2$ in the Hamiltonian; secondly, as we will see in Section \ref{pathologies1}, the St\"uckelberg field satisfies exactly the same equation and thus can be identified with the broken gauge degree of freedom $T(t,x)$.

Observe that the frequency of the scalar mode is $\omega=0$, and thus this is more a zero mode than a propagating particle mode. Nonetheless, this is the physical degree of freedom of our Hamiltonian system. We will postpone the interpretation of this result for the moment and discuss it in Section \ref{conclusion}.

\section{The canonical formalism -- the full theory}
\label{canfor}
We shall now consider the Hamiltonian formulation of the full $n$-DBI gravity theory and confirm the existence of one extra degree of freedom. The $n$-DBI Lagrangian, $L_{\rm nDBI}$, is given by equation (\ref{DBIgravity}).  The square root makes the analysis cumbersome, and we find it more convenient to work in the linearised form with the auxiliary field $e$ discussed in Section \ref{auxfield}:
\bequ
L^{e}_{\text{nDBI}} =  - \frac{1}{\kappa}\int d^{3}x \sqrt{-g}\, e\left( R+K_{ij}K^{ij}-K^2-2N^{-1}\Delta N - 2G_{N} \Lambda_{C}(e)  \right) \ , \label{d1}
\eequ
where $\kappa \equiv 16 \pi G_{N}$. 
In passing to the Hamiltonian formalism, we set the notation for the canonical conjugate momenta as
\begin{equation}
L^{e}_{\text{nDBI}} \left( ( h_{ij}, \dot{h}_{ij} ), ( N, \dot{N} ) , ( N_{i}, \dot{N_{i}} ) , (e ,  \dot{e}) \right) \rightarrow H^{e}_{\text{nDBI}} \left( ( h_{ij}, p^{ij} ) , ( N , p_{N} )  , ( N_{i},  p_{\vec{N}}^{i} )  , (e , p_{e}) \right) \ . \label{d3}
\end{equation}
However, the time derivatives of $N$ and $N_{i}$ are absent as in GR, and so is the time derivative of $e$. Thus we have the primary constraints,
\begin{equation}
\Phi_{1} \equiv p_{N} =0, \ \ \ \Phi^{i}_{2} \equiv p^{i}_{\vec{N}} = 0, \ \ \ \Phi_{3} \equiv p_{e} = 0 \ . \label{d4}
\end{equation}
Denoting the Lagrangian density  by $\mathcal{L}$, the Hamiltonian density is given by
\begin{align}
\mathcal{H}^{e(0)}_{\text{nDBI}}  & \equiv p^{ij}\dot{h}_{ij} - \mathcal{L}^{e}_{\text{nDBI}} \notag \\ &= \sqrt{h} N_{j} \left(- \frac{2}{\sqrt{h}} \nabla_{i}p^{ij} \right)  + \frac{\sqrt{h}N}{\kappa}  \biggl[ - \frac{  \kappa ^{2} }{eh} \left( p^{ij}p_{ij} - \frac{1}{2}p^{2} \right) + e \left( R - 2 G_{N} \Lambda_{C}(e) \right) \biggl] \notag \\
& \ \ \ - \frac{2}{\kappa} \sqrt{h} \left( e \Delta N \right)  \ ,  \label{d5}
\end{align}
where
\begin{equation}
p^{ij} \equiv \frac{\delta L^e_{\text{nDBI}}}{\delta \dot{h}_{ij}} =- \frac{\sqrt{h}  }{\kappa}(K^{ij}-h^{ij}K)   e \ . \label{d6}
\end{equation}
The time flow of the constraints are generated by the extended Hamiltonian density
\begin{equation}
\mathcal{H}^{e(1)}_{\text{DBI}} = \mathcal{H}^{e(0)}_{\text{DBI}} + \lambda_{1} \Phi_{1} + \lambda_{2i}\Phi^{i}_{2} + \lambda_{3} \Phi_{3}\ ,
\end{equation} 
where $\lambda_{1}$, $\lambda_{2i}$ and $\lambda_{3}$ are the Lagrange multipliers. 
Thus the primary constraints evolve in time as
 \begin{align}
\dot{\Phi}_{1}(x) &= \int d^{3}y\{   \mathcal{H}^{e(0)}_{\text{DBI}}(y)  ,\Phi_{1}(x) \} + \sum_{a = 1,3} \int d^{3}y \{   \Phi_{a}(y) , \Phi_{1} (x)  \} \lambda_{a} + \int d^{3}y \{  \Phi^{i}_{2}(y)   ,\Phi_{1} (x) \}\lambda_{2i} \notag \\
&=  - \frac{\sqrt{h}}{\kappa}  \biggl[ - \frac{ \kappa^{2} }{eh} \left( p^{ij}p_{ij} - \frac{1}{2}p^{2} \right) + e \left( R - 2 G_{N} \Lambda_{C}(e) \right) - 2 \Delta e \biggl] \equiv \Phi_{4}(x) \ , \label{d7}
\end{align}
\begin{align}
\dot{\Phi}^{i}_{2}(x) &= \int d^{3}y\{  \mathcal{H}^{e(0)}_{\text{DBI}}(y) ,   \Phi^{i}_{2}(x)  \} + \sum_{a = 1,3} \int d^{3}y \{   \Phi_{a}(y) ,  \Phi^{i}_{2}(x)  \} \lambda_{a} + \int d^{3}y \{  \Phi^{j}_{2}(y) , \Phi^{i}_{2} (x)  \}\lambda_{2j} \notag \\
&=  2 \nabla_{j}p^{ij} \equiv \Phi^{i}_{5}(x) \ , \label{d8}
\end{align} 
 \begin{align}
\dot{\Phi}_{3}(x) &= \int d^{3}y\{   \mathcal{H}^{e(0)}_{\text{DBI}}(y)  ,\Phi_{3}(x) \} + \sum_{a = 1,3} \int d^{3}y \{   \Phi_{a}(y) , \Phi_{3} (x)  \} \lambda_{a} + \int d^{3}y \{  \Phi^{i}_{2}(y)   ,\Phi_{3} (x) \}\lambda_{2i} \notag \\
&=   \frac{\sqrt{h}N}{\kappa e}  \biggl[ - \frac{ \kappa^{2} }{eh} \left( p^{ij}p_{ij} - \frac{1}{2}p^{2} \right) - e \left( R - 2 G_{N} \Lambda_{C}(e) \right) - \frac{12 \lambda}{G_{N}} \left( q  - \frac{1}{e} \right) \biggl] \notag \\ 
& \ \ \ + \frac{2\sqrt{h}}{\kappa} \Delta N  \equiv \Phi_{6}(x) \ . \label{d9}
\end{align}
Therefore, in addition to the primary constraints (\ref{d4}), we have the secondary constraints
\begin{align}
 \Phi_{4} = \dot{p}_{N}\approx 0\ , \ \ \   \Phi^{i}_{5} = \dot{p}^{i}_{\vec{N}}  \approx 0 \ , \ \ \  \Phi_{6} = \dot{p}_{e} \approx 0 \ .  \label{d10}
\end{align}
Finally, the time flows of the secondary constraints do not yield any further constraints, as shown in Appendix \ref{compconstraints}. 
%

\subsection{Constraints and counting degrees of freedom}
\label{seccount}
It is easy to understand the physical meaning of the above constraints. Firstly, we can solve $\Phi_{6}=0$ for $e$ and obtain
\begin{align}
e = \pm  \left( 1 + \frac{G_{N}}{6 \lambda} \mathcal{R}    \right)^{-1/2}\ , \label{11.0}
\end{align}
where we defined ${\cal R}\equiv \mbox{}^{(4)}\! R+\mathcal{K}$. 
Choosing the positive sign and plugging it into $\Phi_{4}$, we find
\begin{align}
\Phi_{4}&=  - \frac{\kappa}{2} \biggl[ - \frac{3 \lambda }{4\pi G^{2}_{N}}q+  \frac{1}{ \sqrt{h} }   \sqrt{ A(h,N)B(h,p)  } +  \frac{1}{\sqrt{h}}\frac{\Delta N}{N} \sqrt{\frac{B(h,p)}{A(h,N)}}  -  \Delta \left( \frac{1}{\sqrt{h}} \sqrt{\frac{B(h,p)}{A(h,N)}} \right) \biggl] \ , \label{d11.1}
\end{align} 
where
\begin{equation}
A(h,N) =  \frac{6\lambda }{G_{N}}  + R -2N^{-1}\Delta N, \ \ \ B(h,p) = 2p^{2} -4p^{ij}p_{ij} +              \frac{3\lambda h}{32 \pi^{2}G^{3}_{N}} \ . \label{18}
\end{equation} 
In terms of the Lagrangian variables, the constraint $\Phi_4=0$ yields
\begin{equation}
0 = \frac{1+\frac{G_{N}}{6\lambda}  ( R-N^{-1}\Delta  N )  }{\sqrt{1+ \frac{G_{N}}{6 \lambda}\mathcal{R} }} -q-\frac{G_{N}}{6 \lambda} \Delta \left(      1+ \frac{    G_{N}   }{      6 \lambda   } \mathcal{R}       \right)^{-1/2} \  .\label{24}
\end{equation}
This is nothing but the Hamiltonian constraint in \cite{Herdeiro:2011im}. 
Similarly, plugging (\ref{11.0}) into (\ref{d6}), the constraints $\Phi_5^i=0$ yield
\begin{align}
\nabla_{j}\left(    \frac{K^{ji} - h^{ji}K}{\sqrt{    1+ \frac{G_{N}}{6 \lambda} \mathcal{R}     }}    \right) = 0 \ . \label{23}
\end{align}  
These are the momentum constraints in \cite{Herdeiro:2011im}. We, however, note that it is more appropriate to regard the following  linear combination
\begin{equation}
\tilde{\Phi}_{5j} = \Phi_{5j} - \Phi_{1} \partial_{j}N - \Phi_{3} \partial_{j}e \ ,
\end{equation} 
as the momentum constraints. This is because these are the constraints that generate the spatial diffeomorphisms for the phase-space variables, $h_{ij}$, $p^{ij}$, $N$ and $e$, rather than $\Phi_{5j}$ (see Appendix \ref{compconstraints}). 

By an explicit computation, one can show that $\Phi_{2j}$ and $\tilde{\Phi}_{5j}$ are first-class constraints, forming the constraint algebra (\ref{CA1})--(\ref{CA6}), and the rest are second class (see Appendix \ref{compconstraints} for details).  
Hence, the constraints are classified as
\begin{align}
& \# (\text{phase-space variables}) = \#  (h_{ij}, p^{ij} , N , p_{N} , N_{i} , p^{i}_{\vec{N}} , e , p_{e}  ) = 2(6+1+3+1) = 22\ , \notag \\
&  \# (\text{$2$nd-class constraints}) = \# ( \Phi_{1}, \Phi_{3} , \Phi_{4} , \Phi_{6}) = 1+1+1+1 = 4\ , \notag \\
& \# (\text{1st-class constraints}) = \# (\Phi_{2j} , \tilde{\Phi}_{5j}) = 3 + 3 = 6\ . \notag
\end{align}
Consequently, the number of physical degrees of freedom (DOF) in $n$-DBI gravity reads
\begin{equation}
\text{DOF of graviton} = \frac{1}{2} \left( 22 - 4 - 2 \times 6 \right) = 2+1 \ . \label{b2}
\end{equation} 
Indeed, as advertised, we find an extra degree of freedom as compared to GR, which, as seen in the previous section, is a scalar mode, hence a \textit{scalar graviton}.

\section{Potential Pathologies}
\label{pathologies}
We shall now address some potential pathologies associated with the existence of the scalar graviton, which are known to afflict HL gravity.

\subsection{Vanishing lapse}
One of the most serious problems of HL gravity is the absence of dynamics discussed in \cite{Henneaux:2009zb}. The problem is that the time flow of the Hamiltonian constraint yields an independent constraint on the lapse, which, as it turns out, requires the lapse to identically vanish in asymptotically flat space-times, implying that HL gravity does not have any dynamics. The authors of \cite{Henneaux:2009zb} emphasise that this problem is intimately related to the fact that in HL gravity the Hamiltonian constraint is second class.

In $n$-DBI gravity, the Hamiltonian constraint is also second class. This raises the concern that this model might also lack dynamics. 
However, this problem is evaded in a similar way as the consistent extension of HL gravity does. Namely, the time flow of the Hamiltonian constraint,  given by (\ref{ap1}) in the full theory, does not yield an additional constraint unlike HL gravity; rather it determines the Lagrange multiplier $\lambda_3$, since the Hamiltonian constraint $\Phi_4$ depends on the auxiliary field $e$ (and implicitly on the lapse $N$ through $e$ by solving $\Phi_6=0$) and thus $\{\Phi_3, \Phi_4\}\ne 0$.
This stems from the {\it nonlinear} lapse dependence in the $n$-DBI action and can be seen more clearly in the linearised theory in Section \ref{gravexists}:\footnote{In the full theory, the nonlinear lapse dependence is somewhat obscured by the introduction of the auxiliary field $e$ which linearises the lapse dependence.} the time flow of the Hamiltonian constraint $\Phi_4$ is generated by its commutator with the Hamiltonian density (\ref{hamper3}). Because of the lapse term $\Delta^2n$ in the Hamiltonian constraint descended from the nonlinear lapse term $(\Delta n)^2\sim (N^{-1}\Delta N)^2$ in the action, the coefficient $\{\Phi_1, \Phi_4\}$ of the Lagrange multiplier $\lambda_1$ is non-vanishing. Thus, such time flow determines $\lambda_1$ rather than imposing an extra constraint on the lapse. 

To summarize, the nonlinear lapse dependence intrinsic to $n$-DBI gravity rescues the model from the absence of dynamics.

\subsection{Short distance instability}
\label{pathologies1}
In \cite{Blas:2009yd} it was found that the scalar graviton in HL gravity develops an exponential time growth at short distances in generic space-times in the linearised approximation. This suggests the presence of a universal short distance instability in HL gravity. Such problem, however, should be considered with some caution. The exponentially growing mode involves the lapse; but the lapse in asymptotically flat space-times is forced to vanish everywhere \cite{Henneaux:2009zb}, as mentioned in the previous section. If the lapse must collapse everywhere, there would be no way to develop an exponentially growing mode involving the lapse. Thus there appears to be a contradiction between these two issues. This might imply that the unstable scalar mode found in \cite{Blas:2009yd} fails to obey the boundary condition at spatial infinity, when extended to long distances. In any case, as we have just shown, the lapse does not vanish in our model and so it is sensible and important to study whether the scalar mode leads to any sort of instability or not.

We shall now show that the analysis that reveals an exponential time growth at short scales in HL gravity, does not allow the same conclusion when applied to the scalar graviton in $n$-DBI. 
We begin by recasting the equations of motion of $n$-DBI gravity:
\begin{align}
&(\partial_t-\mathcal{L}_{N})h_{ij}-2NK_{ij}=0 \ ,\label{EQmetric}\\
&R+K^2-K_{ij}K^{ij}=2G_N\Lambda_C(e)+2e^{-1}\Delta e\ ,\label{EQham}\\
&\nabla^j\left(K_{ij}-h_{ij}K\right)=-e^{-1}\nabla^je\left(K_{ij}-h_{ij}K\right)\ ,\label{EQmom}\\
&(\partial_t-\mathcal{L}_{N})\left(K^{ij}-h^{ij}K\right)+N\left(R^{ij}-KK^{ij}+2K^{il}K^{j}_{\phantom{j}l}-h^{ij}K_{mn}K^{mn}\right)-\nabla^i\nabla^j N+h^{ij}\Delta N\nonumber\\
&\hspace{2.5cm}=-e^{-1}(\partial_t-\mathcal{L}_{N})e\left(K^{ij}-h^{ij}K\right)+e^{-1}N\left(\nabla^i\nabla^j-h^{ij}(\nabla_l \ln N)\nabla^l\right)e\label{EQevol}\ ,
\end{align}
where
$e=\left(1+\frac{G_N}{6\lambda}{\cal R}\right)^{-\half}$ and the cosmological constant $\Lambda_C(e)={3\lambda\over G_N^2}\left({2q\over e}-1-{1\over e^{2}}\right)$.
The first equation defines the extrinsic curvature. The second and third equations are the Hamiltonian and momentum constraints, respectively. The last equation is the evolution equation \cite{Herdeiro:2011im}.
In this form, the R.H.S. of (\ref{EQham})--(\ref{EQevol}) represent the modification from GR. 
In fact, when the auxiliary field $e$ is constant, the R.H.S. vanish except for the cosmological constant in the Hamiltonian constraint (\ref{EQham}), and these equations become those of GR, as remarked in Section \ref{auxfield}. 
As we have seen in (\ref{scalargravitoneq}) in Section \ref{312}, the time-derivative of the Hamiltonian constraint succinctly captures the scalar mode dynamics, and it can be written as
\begin{equation}
\Delta\dot{e}-KN\Delta e-\left(N\del^iK+2K\del^iN\right)\del_ie-N^{-1}\Delta N\dot{e}=0\ .\label{EQscalar}
\end{equation}

We now consider small perturbations around a generic background in the gauge $N^i=0$:
\begin{eqnarray}
h_{ij}&=&\bar{h}_{ij}+\gamma_{ij}\ ,\\
K^{ij}&=&\bar{K}^{ij}+\kappa^{ij}\ ,\\
N&=&\bar{N}+n\ .
\end{eqnarray}
The details of the following analysis can be found in Appendix \ref{perturbation_eq_motion}.
We assume that the background space-time varies over the characteristic scale $L$; $\bar{R}_{ij}\sim1/L^2$, $\bar{K}_{ij}\sim1/L$, $\bar{\mathcal{R}}\sim1/L^2$, $\del\ln\bar{N}\sim 1/L$, and $\bar{e}\sim1+O(1/L^2)$. 
Since we are interested in a potential universal short distance instability, 
we focus on scales much shorter than $L$;
\bequ
\omega,p \gg1/L \ , \label{assumption}
\eequ
where the space-time is nearly flat. We can then neglect terms with space-time derivatives of the background fields relative to those of the perturbed fields.
Assuming the Fourier decompositions
\begin{equation}
\gamma_{ij}, \kappa^{ij}, n \sim e^{-i\omega t+ip\cdot x}\ ,
\end{equation}
the perturbations of (\ref{EQmetric})--(\ref{EQevol}) and $(\ref{EQscalar})$ yield (\ref{pert_metric})--(\ref{pert_evol}) and (\ref{pert_scalar}), respectively. 
We then find that for the scalar mode to be present
\begin{equation}
\omega\sim \bar{N}p\sim  i\left(\bar{N}\bar{K}+\del\bar{N}\right)\sim\frac{i}{L} \ .
\label{disrel}
\end{equation}
At first sight, it might seem to imply the existence of an exponentially growing mode, suggesting an instability. This relation, however, violates the validity of our approximation (\ref{assumption}) and thus cannot be satisfied. We conclude that this analysis does not reveal a potential universal short distance instability as discussed in the HL case \cite{Blas:2009yd}.
Note that this also implies that there is no scalar mode of the type $(n, \gamma)\sim \left(n(\omega, p), \gamma(\omega, p)\right)e^{i\omega t +ip\cdot x}$ at short distances.\footnote{As we have seen in section \ref{312}, the scalar mode {\it does exist},  but it has $\omega=0$. This can also be seen in the St\"uckelberg formalism, cf. (\ref{phi3}). The scalar mode is neither exponentially growing nor oscillating to linear order. We will discuss more about the scalar mode in Section \ref{conclusion}.}

The same conclusion can be reached by an alternative analysis using the St\"uckelberg field $\phi$. We leave the details of the analysis in Appendix \ref{perturbation_stuckelberg}. The equation of motion for $\phi$ can be obtained from the action (\ref{DBIgravitycovariant}). The preferred choice of time-like vector ${\bf n}=(-N,0,0,0)$ corresponds to $\phi=t$. Indeed, the equation of motion (\ref{eq_scalar}) for $\phi$ reduces,  when $\phi=t$, to the time derivative of the Hamiltonian constraint (\ref{EQscalar}).

We now consider perturbations in unitary gauge 
$\phi=t+\varphi$ and $N^i=0$,
expanding (\ref{EQscalar}) to linear order in $\varphi$.  To leading order, we find 
\begin{equation}
\Delta^2\ddot{\varphi}-(\bar{N}\bar{K}-\del_t\ln \bar{N})\Delta^2\dot{\varphi}
+2\del_i\ln \bar{N}\del^i\Delta\ddot{\varphi}=0\ . \label{phi3}
\end{equation}
This indeed yields (\ref{disrel}) found in the previous analysis.

We conclude by remarking that, as advertised, $\varphi(t,x)$ can be identified with the broken gauge parameter $T(t,x)$. Since in flat space-time (\ref{phi3}) yields $\Delta^2\ddot{\varphi}=0$, we obtain precisely the scalar graviton equation  (\ref{scalargravitoneq}) whose origin can be traced to the nonlinear lapse term (observe also that (\ref{phi3}) is the time derivative of the linearised Hamiltonian constraint). This provides a consistency check of our computations.

\subsection{Strong coupling}
\label{strongcoupling}


Naively, $n$-DBI gravity admits the GR limit by tuning the parameter $\lambda\to\infty$ and $q\to 1$ with $\lambda(q-1)$ fixed finite. However, as is well-known, it is subtle and far from obvious whether the extra scalar mode actually decouples in the IR in the putative GR limit.
A famous example is the vDVZ discontinuity of massive gravity \cite{vanDam:1970vg} which can be attributed to the strongly coupled nature of a scalar mode, as elucidated in \cite{ArkaniHamed:2002sp}. Massive gravity has three extra modes; a transverse spin one and a scalar mode. As it turned out, the scalar mode remains strongly coupled in the massless limit and thus the GR limit does not exist. 

The original form of HL gravity has a similar strong coupling problem \cite{Blas:2009yd, Charmousis:2009tc}; there is an energy scale $\Lambda_s$ above which the scalar mode self-coupling becomes strong. For HL gravity to flow to GR in the IR, the strong coupling scale $\Lambda_s$ needs to be sufficiently high so that the coupling becomes weak and virtually decouples in the IR. However, as it turned out, the naive estimate of the strong coupling scale yields a value too low, $\Lambda_s=\sqrt{\lambda_{\rm HL}-1}\,M_P$, where $\lambda_{\rm HL}$ is an anisotropy parameter and the coefficient of the $K^2$ term;
this is most manifest in the St\"uckelberg field formalism. The part of the St\"uckelberg field action, coming from the $K^2$ term, is proportional to $(\lambda_{\rm HL}-1)M_P^2$. This is the most relevant part of the action in the IR. This implies that the coupling is $g=1/(\sqrt{\lambda_{\rm HL}-1}\,M_P)$ and thus irrelevant. Although the coupling becomes weak in the IR, the strong coupling scale $\Lambda_s$ is too low, approaching zero in the putative GR limit $\lambda_{\rm HL}=1$.
In fact, as argued in \cite{Blas:2009yd}, the situation appears to be even worse. The quadratic action in flat space lacks time derivatives, while there are time derivative interactions. This may imply that the high frequency modes are always strongly coupled. Indeed, one can refine the estimate of the strong coupling scale by making the space-time slightly curved (with the characteristic length scale $L$) which introduces a quadratic term with a time derivative. Then the refined strong coupling scale becomes anisotropic and is estimated to be $\Lambda_{\omega}=L^{-1/4}\Lambda_s^{3/4}$ and  $\Lambda_{p}=L^{-3/4}\Lambda_s^{1/4}$  \cite{Blas:2009yd}. In the flat space limit $L\to\infty$, these are in fact zero.

In $n$-DBI, the situation turns out to be more subtle. The St\"uckelberg field action in flat space  in unitary gauge yields
\begin{align}
S_3={1\over 192\pi \lambda}\int d^4x
\left[\half(\Delta\dot{\varphi})^2-\left(\dot{\varphi}\Delta\ddot{\varphi}
-\del_i\varphi\Delta\del_i\dot{\varphi}\right)\Delta\varphi
+{5\over 12}{(\Delta\dot{\varphi})^3\over \lambda M_P^2}\right] \ ,
\end{align}
 to cubic order, as shown in Appendix \ref{perturbation_stuckelberg2}. The quadratic action is invariant under the scaling $E\to sE$, $t\to s^{-1}t$, $x\to s^{-1} x$, and $\varphi\to s^{-1}\varphi$.
First note that the last interaction is irrelevant as compared to the middle ones. The higher order interactions have a similar structure. So the most ``relevant" terms are of the middle type and actually (classically) marginal, with coupling $\lambda^{\frac{n-2}{2}}$, for the $n$-th order in $\varphi$.\footnote{Most easily seen by redefining $\phi\rightarrow\sqrt{\lambda}\phi$.}
However, due to quantum effects, these operators may become either relevant or irrelevant. For the cubic terms with coupling $g=\sqrt{\lambda}$, the beta function at $1$-loop is of the form $\beta(g)=-cg^3$, from which one obtains\footnote{Here we assume that the IR divergences can be properly regularized.}
\begin{equation}
g^2=g_0^2\left(1+2cg_0^2\ln\frac{\Lambda}{M_P}\right)^{-1},
\end{equation}
where $\Lambda$ is the energy scale and the UV cutoff scale is $M_P$. Therefore, the strong coupling scale is
\begin{equation}
\Lambda_s\sim M_P\exp\left(-{\frac{1}{2c\lambda}}\right).
\end{equation}
This implies that in the putative GR limit $\lambda\rightarrow\infty$, the strong coupling scale is about the Planck scale and thus sufficiently high. 
However, for this coupling to be (marginally) irrelevant, the constant $c$ must be negative. Given the fact that this is a scalar field theory and the higher derivative nature of the operators, it seems plausible that this is actually the case.
Therefore, $n$-DBI gravity is likely to be free from the strong coupling problem.

\section{Discussions}
\label{conclusion}
The first main result of this paper is to have established that $n$-DBI gravity has three degrees of freedom: the usual two tensorial modes of the graviton plus a scalar mode which, however, is not a propagating particle with a definite dispersion relation. The second result is that the arguments for three pathological features of the original formulation of HL gravity do not carry over to $n$-DBI gravity; these arguments concern the absence of dynamics, a universal short distance instability, and a strong coupling issue. The reason why $n$-DBI looks healthier than HL gravity, in these respects, has a common ground: the nonlinear dependence on (spatial derivatives of) the lapse in the action.

The fact that the same arguments for the pathologies in HL gravity do not directly apply to $n$-DBI gravity does not mean, of course, that the latter is free of pathologies. Although the arguments presented herein are reassuring, in order to establish an healthy behaviour, $n$-DBI gravity must be further studied. In particular, we found that the scalar mode grows linearly in time in the linear approximation and is thus at the threshold between stability (oscillation) and instability (exponential growth). In other words, the scalar mode can be either marginally stable or unstable.\footnote{As an illustration, consider a simple mechanical model. For canonical kinetic term, a linear time growth is the behaviour seen in a flat potential. The flat potential does not indicate an instability and is marginally stable. However, any potentials which are flat to quadratic order all lead to the linear time growth for perturbations in the linearised approximation. The simplest examples are cubic and (convex) quartic potentials. The former is clearly unstable (inverse-squared blow-up), whereas the latter is stable (oscillation).} The (in)stability depends on the details of the scalar mode interactions. We thus need to extend the analysis performed in section \ref{312} beyond linear order to establish the (in)stability of the scalar mode. 

\begin{figure}
\centering
\hspace{-1.cm}
\includegraphics[height=2in]{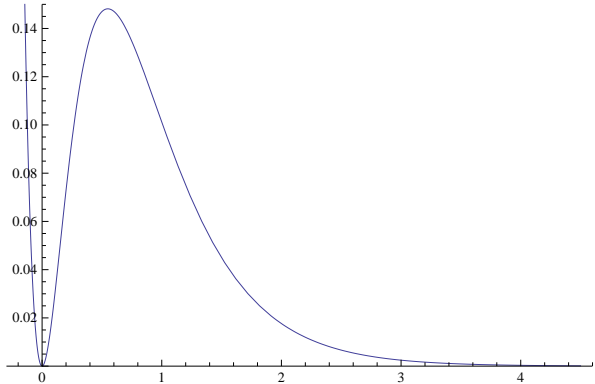}
\caption{The potential $V(\chi)$ defined in (\ref{potential}) with $q=1$} 
\label{v}
\end{figure}

To study the effect of nonlinear interactions, we find it useful to work in the Einstein frame (\ref{Einstein}). Note that the background space-time remains flat under the frame change. A key observation is that, as suggested in Appendix \ref{einstein_frame}, the auxiliary field $\chi$ can essentially be thought of as the time derivative of the scalar mode, {\it i.e.}, $\chi\sim \dot{T}\sim \dot{\varphi}$. In other words, the proof of the stability amounts to showing the stability of the $\chi$ fluctuation. In fact, the equation of motion for $\chi$ in the $N^i=0$ gauge
\begin{align}
\del_t(N^{-2}\dot{\chi})+\Delta\chi-\frac{1}{3}\del_t(N^{-1}K)-\frac{1}{12}V'(\chi)=0\label{psi_eqn}
\end{align}
seems to indicate the stability against small perturbations around $\chi=0$; as is clear, the potential $V(\chi)$ plotted in Fig.\ref{v} is very stable around $\chi=0$.  
Note, however, that the lapse $N$ and the spatial metric $h_{ij}$ also furnish the scalar mode. Moreover, the auxiliary field $\chi$ is further constrained by the other equations of motion. So, strictly speaking, we need to deal with the coupled system of $\chi$, $N$ and $h_{ij}$ (in the $N^i=0$ gauge). There is, however, a shortcut: it would be sufficient to show that the solutions of (\ref{psi_eqn}) for $\chi$ in {\it generic backgrounds} are oscillatory. 

The force in (\ref{psi_eqn}) consists of the strongly attractive force ${1\over 12}N^2V'(\chi)$, the repulsive force $-N^2\Delta\chi=N^2p^2\chi$,  the friction $\del_t\ln N^2\dot{\chi}\sim L^{-1}\dot{\chi}$, and the external force ${1\over 3}N^2\del_t(N^{-1}K)\sim L^{-2}$.
The last three forces may work against  stability. These become more dominant when $\chi$ is smaller. However, the small $\chi$ behaviour is well approximated by the linear perturbation. 
As shown in Appendix \ref{einstein_frame}, $\chi$ is marginally stable to linear order. Meanwhile, for larger $\chi$, the attractive force, being exponential, may quickly become dominant, as $\chi$ increases, and pull it back to smaller $\chi$. Hence it seems plausible that $\chi$ oscillates about $\chi=0$. We therefore conclude that $n$-DBI gravity is likely to be stable against the scalar mode perturbations.

To close, we have considerably improved our understanding of the scalar mode and presented arguments for its healthy behaviour. However, the scalar mode still remains somewhat elusive; it does not appear to behave as a conventional particle with a (non-)relativistic dispersion relation around near-flat space-times. So it is rather unclear what signatures exactly it might leave as an observable. On this score, however, we anticipate that the scalar mode could be the source of scalar perturbations in the Cosmic Microwave Background Radiation, when applied to inflationary cosmology \cite{Herdeiro:2011km}, and we hope to return to this question in the near future.
Finally, we believe that models such as $n$-DBI gravity are interesting learning grounds for deepening our understanding of (non-)relativistic gravity. Needless to say, hurdles are high to self-consistently modify Einstein's elegant theory of general relativity. 

\section*{Acknowledgement}

We would like to thank Marco Sampaio for useful discussions. 
S.H. and Y.S. would like to thank the Niels Bohr Institute and the University of Aveiro for their warm hospitality. F.C. is funded by FCT through the grants SFRH/BD/60272/2009. This work was partially supported by the Grant-in-Aid for Nagoya University Global COE Program (G07), by FCT (Portugal) through the project CERN/FP/116341/2010  and 
PTDC/FIS/116625/2010 and by the Marie Curie Action NRHEP--295189-FP7-PEOPLE-2011-IRSES.


\appendix
\renewcommand{\theequation}{\Alph{section}.\arabic{equation}}

\section{Computations of constraint algebra}
\label{compconstraints}
In this Appendix, we give an explicit computation of the constraint algebra and classification of the constraints presented in Section \ref{canfor}.
To facilitate the computation, we introduce smooth test vector fields, $\xi^{\mu} = (\xi^{0} , \xi^{i})$ and $\eta^{\mu} = (\eta^{0} , \eta^{i})$, which fall off fast enough to 
suppress all the boundary contributions \cite{Khoury:2011ay}. Henceforth, we define the smeared constraints:
\begin{align}
&\hat{\Phi}_{1} (\xi^{0}) = \int d^{3} x \xi^{0} (x) \Phi_{1} (x) \ , \\
&\hat{\Phi}_{2} (\xi^{i}) = \int d^{3} x \xi^{i} (x) \Phi_{2i} (x) \ , \\
& \hat{\Phi}_{3} (\xi^{0}) = \int d^{3} x \xi^{0} (x) \Phi_{3} (x) \ , \\
& \hat{\Phi}_{4}(\xi^{0}) = \int d^{3}x \xi^{0}(x)\Phi_{4}(x) \ ,  \\
& \hat{\Phi}_{5} (\xi^{i}) = \int d^{3} x \xi^{i}(x) \Phi_{5i}(x) \ , \\
& \hat{\Phi}_{6} (\xi^{0}) = \int d^{3} x \xi^{0}(x) \Phi_{6} (x) \ , \\
& \hat{\Phi}^{(G)}_{N}(\xi^{i}) = \int d^{3}x \xi^{i}(x) \Phi^{(G)}_{N i }(x) = \int d^{3}x \xi^{i}(x) ( - \Phi_{1}\partial_{i} N )(x)\ , \label{GN}\\
& \hat{\Phi}^{(G)}_{e}(\xi^{i}) = \int d^{3}x \xi^{i}(x) \Phi^{(G)}_{ei}(x) = \int d^{3}x \xi^{i}(x) ( - \Phi_{3}\partial_{i} e )(x) \ , \\
& \hat{\Phi}^{(G)}_{\vec{N}}(\xi^{i}) = \int d^{3}x \xi^{i}(x) \Phi^{(G)}_{ \vec{N}i}(x) = \int d^{3}x \xi^{i}(x) \biggl[- \left( \Phi^{i}_{2}\nabla_{i}N^{j} + \nabla_{j} \left( \Phi_{2 i} N^{j} \right)   \right) (x) \biggl] \ , \\
& \hat{\Phi}'_{5} (\xi^{i}) = \hat{\Phi}_{5} (\xi^{i}) +  \hat{\Phi}^{(G)}_{N}(\xi^{i}) + \hat{\Phi}^{(G)}_{e}(\xi^{i}) = \int d^{3} x \xi^{i}(x) \tilde{\Phi}_{5i}(x) \ ,\label{Phi5tilde}
\end{align}  
where $\Phi_{1}$, $\Phi_{2i}$, $\Phi_{3}$, $\Phi_{4}$, $\Phi_{5i}$, $\Phi_{6}$ and $\tilde{\Phi}_{5i}$ are the constraints defined in Section \ref{canfor}. 
The idea is to compute the commutators like
\begin{equation}
\{ \hat{\Phi}_{4}(\xi^{0}) , \hat{\Phi}_{5}(\eta^{j}) \} = \int d^{3}y d^{3}x \xi^{0}(y)\eta^{j}(x) \{ \Phi_{4}(y) , \Phi_{5j}(x) \} \ ,
\end{equation}
and read off the algebra of local constraints from the R.H.S. 
 
The basic non-vanishing Poisson's brackets are given by
\begin{align}
& \{ p^{ij} (y) , h_{kl} (x)  \} = \frac{1}{2} ( \delta^{i}_{k} \delta^{j}_{l} + \delta^{i}_{l} \delta^{j}_{k}  ) \delta (y-x) \ , \\
& \{ p_{N} (y) , N(x)  \} = \delta (y-x) \ , \\
& \{ p^{i}_{\vec{N}} (y) , N_{j} (x) \} = \delta^{i}_{j} \delta (y-x) \ , \\
& \{ p_{e} (y) , e(x) \} = \delta (y-x) \ .
\end{align}
To compute the Poisson brakets of constraints, we take the variations of the smeared constraints:
\begin{align}
&  \frac{\delta \hat{\Phi}_{1}(\xi^{0})}{\delta p_{N}} = \xi^{0} \ ,  \\ 
&  \frac{\delta \hat{\Phi}_{2}(\xi^{i})}{\delta p_{\vec{N}i}} = \xi^{i} \ ,  \\ 
&  \frac{\delta \hat{\Phi}_{3}(\xi^{0})}{\delta p_{e}} = \xi^{0} \ ,  \\ 
&  \frac{\delta \hat{\Phi}_{4}(\xi^{0})}{\delta h_{mn}} = \xi^{0} \biggl[ \frac{1}{2}h^{mn}\Phi_{4}  - \frac{\kappa}{e \sqrt{h}}h^{mn}\left( p^{ab}p_{ab} - \frac{1}{2}p^{2} \right) + \frac{2 \kappa}{e\sqrt{h}} \left( p^{ml}p^{n}_{l} - \frac{1}{2}p^{mn}p   \right) \notag \\
 & \ \ \  \ \ \ \ \ \ \ \ \ \ + \frac{e}{\kappa} \sqrt{h} R^{mn} - \frac{\sqrt{h}}{\kappa} (\nabla^{m} \nabla^{n} e)    \biggl] \notag \\
& \ \ \  \ \ \ \ \ \ \ \ \ \  +  \frac{\sqrt{h}}{\kappa} \biggl[ - (\nabla^{m}\nabla^{n} \xi^{0})e + h^{mn} (\nabla^{a} e)(\nabla_{a} \xi^{0}) + h^{mn} e (\Delta \xi^{0})   \biggl] \ , \\
& \frac{\delta \hat{\Phi}_{4}(\xi^{0})}{\delta p^{mn}} = \xi^{0} \biggl[ \frac{2 \kappa}{e \sqrt{h}} \left( p_{mn} - \frac{1}{2}h_{mn}p       \right)  \biggl]\ , \\
&  \frac{\delta \hat{\Phi}_{4}(\xi^{0})}{\delta e} = - \xi^{0} \biggl[  \frac{\kappa }{e^{2}\sqrt{h}} \left( p^{ab}p_{ab} - \frac{1}{2}p^{2}   \right) + \frac{\sqrt{h}}{\kappa} R + \frac{6 \lambda \sqrt{h}}{\kappa G_{N}} \left( 1 - \frac{1}{e^{2}} \right)   \biggl] + \frac{2 \sqrt{h}}{\kappa} \Delta \xi^{0}\ , \\
& \frac{\delta \hat{\Phi}_{5}(\xi^{i})}{\delta h_{mn}} = - (\nabla_{l}\xi^{i})p^{lj}(\delta^{m}_{i}\delta^{n}_{j} + \delta^{m}_{j} \delta^{n}_{i}) + \nabla_{l} (p^{mn} \xi^{l}) = \mathsterling_\xi p^{mn} \ , \\
&\frac{\delta \hat{\Phi}_{5}(\xi^{i})}{\delta p^{mn}} = -h_{li}(\nabla_{j} \xi^{l}) (\delta^{i}_{m}\delta^{j}_{n} + \delta^{i}_{n} \delta^{j}_{m}) = - \mathsterling_\xi h_{mn}  \ , \\
&\frac{\delta \hat{\Phi}_{6}(\xi^{0})}{\delta h_{mn}} = \xi^{0} \biggl[ \frac{1}{2} h^{mn}\Phi_{6} + \frac{N \kappa}{e^{2} \sqrt{h}}h^{mn} \left( p^{ab}p_{ab} - \frac{1}{2}p^{2}   \right) - \frac{2 \kappa N}{e^{2}\sqrt{h}} \left( p^{ml}p^{n}_{l} - \frac{1}{2}p^{mn}p   \right) \notag \\
& \ \ \  \ \ \ \ \ \ \ \ \ \ + \frac{N\sqrt{h}}{\kappa} R^{mn} - \frac{\sqrt{h}}{\kappa} (\nabla^{m} \nabla^{n} N)    \biggl] \notag \\
& \ \ \  \ \ \ \ \ \ \ \ \ \  + \frac{\sqrt{h}}{\kappa} \biggl[ -N (\nabla^{m}\nabla^{n} - h^{mn} \Delta)\xi^{0} + h^{mn}( \nabla^{l} N) (\nabla_{l}\xi^{0})   \biggl]\ , \\
& \frac{\delta \hat{\Phi}_{6}(\xi^{0})}{\delta p^{mn}} = - \xi^{0} \biggl[  \frac{2N \kappa}{e^{2}\sqrt{h}} \left( p_{mn} - \frac{1}{2}h_{mn}p   \right) \biggl] \   , \\
& \frac{\delta \hat{\Phi}_{6}(\xi^{0})}{\delta N} =  - \xi^{0} \biggl[ \frac{\kappa}{e^{2} \sqrt{h}} \left( p^{ab}p_{ab} - \frac{1}{2}p^{2} \right) + \frac{\sqrt{h}}{\kappa} R + \frac{6 \lambda \sqrt{h}}{\kappa G_{N}} \left( 1 - \frac{1}{e^{2}} \right)   \biggl]  + 2 \frac{\sqrt{h}}{\kappa} \Delta (\xi^{0})\ , \\
&\frac{\delta \hat{\Phi}_{6}(\xi^{0})}{\delta e} = \xi^{0} \biggl[ \frac{2 N \kappa}{e^{3} \sqrt{h}} \left( p^{ab}p_{ab} - \frac{1}{2} p^{2}   \right) - \frac{12 \lambda N \sqrt{h}}{\kappa e^{3} G_{N}}   \biggl] \ , \\
& \frac{\delta \hat{\Phi}^{(G)}_{N} (\xi^{i})}{\delta N} =   \partial_{i} ( \xi^{i}  p_{N} ) \ , \\
& \frac{\delta \hat{\Phi}^{(G)}_{N}(\xi^{i})}{\delta p_{N}} = -  \xi^{i} \partial_{i} N = - \mathsterling_\xi N \ , \\
& \frac{\delta \hat{\Phi}^{(G)}_{e} (\xi^{i})}{\delta e} =   \partial_{i} ( \xi^{i}  p_{e} ) \ , \\
& \frac{\delta \hat{\Phi}^{(G)}_{e}(\xi^{i})}{\delta p_{e}} = -  \xi^{i} \partial_{i} e = -\mathsterling_\xi e \ , \\
& \frac{\delta \hat{\Phi}^{(G)}_{\vec{N}}(\xi^{i})}{ \delta N^{j}} = \nabla_{i} (\xi^{i} p_{\vec{N} j } ) + (\nabla_{j} \xi^{i}) p_{\vec{N}i} \ , \\
& \frac{\delta \hat{\Phi}^{(G)}_{\vec{N}}(\xi^{i})}{\delta p_{\vec{N}}} = -  \xi^{i} \nabla_{i} N^{j} + N^{i} \nabla_{i} \xi^{j} = - \mathsterling_\xi N^{j} \ .  
\end{align}
Using these, it is tedious but straightforward to compute the constraint algebra:
\begin{align}
& \{ \Phi_{1}(y) , \Phi_{6} (x)   \} = -  \biggl[ \frac{\kappa}{e^{2} \sqrt{h}} \left( p^{ab}p_{ab} - \frac{1}{2}p^{2} \right) + \frac{\sqrt{h}}{\kappa} R + \frac{6 \lambda \sqrt{h}}{\kappa G_{N}} \left( 1 - \frac{1}{e^{2}} \right)   \biggl]  \delta (y-x) \notag \\
& \ \ \  \ \ \ \ \ \ \ \ \ \ \ \ \ \ \ \ \ \ + 2\frac{ \sqrt{h} }{ \kappa} \Delta  \delta (y-x) \ , \\
& \{ \Phi_{1} (y) , \Phi^{(G)}_{N j}(x)   \} = \Phi_{1} (x) \partial_{y^{j}} \delta (y-x) \ , \\
& \{ \Phi_{3} (y) , \Phi_{4} (x) \} =  -  \biggl[  \frac{\kappa }{e^{2}\sqrt{h}} \left( p^{ab}p_{ab} - \frac{1}{2}p^{2}   \right) + \frac{\sqrt{h}}{\kappa} R + \frac{6 \lambda \sqrt{h}}{\kappa G_{N}} \left( 1 - \frac{1}{e^{2}} \right)   \biggl] \delta (y-x) \notag \\
&\ \ \  \ \ \ \ \ \ \ \ \ \ \ \ \ \ \ \ \ \   + \frac{2 \sqrt{h}}{\kappa} \Delta \delta (y-x) \ , \\
& \{ \Phi_{3} (y) , \Phi_{6} (x) \} = \biggl[ \frac{2 N \kappa}{e^{3} \sqrt{h}} \left( p^{ab}p_{ab} - \frac{1}{2} p^{2}   \right) - \frac{12 \lambda N \sqrt{h}}{\kappa e^{3} G_{N}}   \biggl] \delta (y-x) \ , \\
& \{ \Phi_{3} (y) , \Phi^{(G)}_{ ej }(x)   \} = \Phi_{3} (x) \partial_{y^{j}} \delta (y-x) \ , \\
& \{ \Phi_{4}(y) , \Phi_{4}(x)   \} = \Phi^{k}_{5}(y) \partial_{y^{k}} \delta (y-x) - \Phi^{k}_{5} (x) \partial_{x^{k}} \delta (y - x) \notag \\
& \ \ \  \ \ \ \ \ \ \ \ \ \ \ \ \ \ \ \ \ \  - \frac{p}{e} \partial_{y^{k}}e \partial_{y_{k}} \delta (y-x) + \frac{p}{e} \partial_{x^{k}}e \partial_{x_{k}} \delta (y - x) \ , \\
&\{ \Phi_{4} (y) , \Phi_{5j} (x)   \} = \Phi_{4} (x) \partial_{y^{j}} \delta (y-x) - \frac{\partial_{x^{j}} e  }{N} \Phi_{6}(x) \delta (y-x) - 2 \frac{\sqrt{h}}{\kappa} \partial_{x^{j}}e \left( \Delta - \frac{\Delta N}{N} \right) \delta (y-x) \ ,  \\
& \{ \Phi_{4} (y) , \Phi_{6} (x) \} =  \frac{2N p_{mn}}{e}   \biggl[ 2R^{mn} - \left( \frac{\nabla^{m}\nabla^{m} e}{e} + \frac{\nabla^{m} \nabla^{n} N}{N}   \right) \biggl] \delta (y-x) \notag \\
& \ \ \  \ \ \ \ \ \ \ \ \ \ \ \ \ \ \ \ \ \ -  \frac{Np}{e}    \biggl[ R - \frac{6 \lambda}{G_{N}} \left( 1 - \frac{q}{e}   \right)   \biggl] \delta (y-x) \notag \\
& \ \ \  \ \ \ \ \ \ \ \ \ \ \ \ \ \ \ \ \ \ - \frac{2Np_{mn} (y)}{e } \nabla^{m}_{(y)} \partial_{y_{n}} \delta (y-x) - \frac{2Np_{mn} (x)}{e}  \nabla^{m}_{(x)} \partial_{x_{n}} \delta (y-x) \notag \\
& \ \ \  \ \ \ \ \ \ \ \ \ \ \ \ \ \ \ \ \ \ - \left( \frac{Np}{e} \right) \frac{\partial_{y^{m}} N }{N} \partial_{y_{m}} \delta (y-x) - \left( \frac{Np}{e}   \right) \frac{\partial_{x^{m}}e}{e} \partial_{x_{m}} \delta (y-x) \ , \\
& \{ \Phi_{4}(y) , \Phi^{(G)}_{ej} (x) \} =  \frac{\partial_{x^{j}} e  }{N} \Phi_{6}(x) \delta (y-x) + 2 \frac{\sqrt{h}}{\kappa} \partial_{x^{j}}e \left( \Delta - \frac{\Delta N}{N} \right) \delta (y-x) \ , \\  
& \{ \Phi_{5j}(y) , \Phi_{5i}(x) \} = \Phi_{5j}(x)\partial_{y^{i}}\delta (y-x) - \Phi_{5i}(y)\partial_{x^{j}}\delta (y-x) \ , \\
& \{ \Phi_{6}(y) , \Phi_{5j}(x)   \} = \Phi_{6} (x) \partial_{y^{j}} \delta (y-x) - \frac{\partial_{x^{j}} N  }{N} \Phi_{6}(x) \delta (y-x) - 2 \frac{\sqrt{h}}{\kappa} \partial_{x^{j}}N \left( \Delta - \frac{\Delta N}{N} \right) \delta (y-x)  \notag \\
&  \ \ \  \ \ \ \ \ \ \ \ \ \ \ \ \ \ \ \ \ \ - \frac{2 N \sqrt{h}}{\kappa}\partial_{y^{j}}e \left(  \frac{B(h,p)}{e^{3}} \right) \delta (y-x) \ , \\ 
& \{ \Phi_{6} (y) , \Phi_{6} (x)   \} = - \partial_{y_{m}} \left( \frac{2N^{2}}{e^{2}}   \right) p_{mn}(y) \partial_{y_{n}} \delta (y-x) + \partial_{x_{m}} \left( \frac{2N^{2}}{e^{2}}   \right) p_{mn} (x) \partial_{x_{n}} \delta (y-x) \notag \\
&  \ \ \  \ \ \ \ \ \ \ \ \ \ \ \ \ \ \ \ \ \ - \left( \frac{N}{e}   \right)^{2} \Phi_{5n}(y) \partial_{y_{n}} \delta (y-x) + \left( \frac{N}{e}   \right)^{2} \Phi_{5n}(x) \partial_{x_{n}} \delta (y-x) \notag \\
& \ \ \  \ \ \ \ \ \ \ \ \ \ \ \ \ \ \ \ \ \ + \frac{Np \partial_{y^{m}} N }{e^{2}} \partial_{y_{m}} \delta (y-x) - \frac{Np \partial_{x^{m}} N}{e^{2}} \partial_{x_{m}} \delta (y-x) \ , \\
& \{ \Phi_{6}(y) , \Phi^{(G)}_{Nj} (x) \} =  \frac{\partial_{x^{j}} N  }{N} \Phi_{6}(x) \delta (y-x) + 2 \frac{\sqrt{h}}{\kappa} \partial_{x^{j}}N \left( \Delta - \frac{\Delta N}{N} \right) \delta (y-x) \ , \\
& \{ \Phi_{6} (y) , \Phi^{(G)}_{ej} (x)   \} =  \frac{2 N \sqrt{h}}{\kappa}\partial_{y^{j}}e    \left(  \frac{B(h,p)}{e^{3}} \right)  \delta (y-x) \  , \\
& \{ \Phi^{(G)}_{Nj} (y) , \Phi^{(G)}_{Ni} (x)   \} = \Phi^{(G)}_{Nj}(x) \partial_{y^{i}} \delta (y-x) - \Phi^{(G)}_{Ni} (y) \partial_{x^{j}} \delta (y-x) \ , \\
&  \{ \Phi^{(G)}_{ej} (y) , \Phi^{(G)}_{ei} (x)   \} = \Phi^{(G)}_{ej}(x) \partial_{y^{i}} \delta (y-x) - \Phi^{(G)}_{ei} (y) \partial_{x^{j}} \delta (y-x) \ , \\
& \{ \Phi_{1} (y), \Phi_{2j} (x)  \} =   \{ \Phi_{2i} (y), \Phi_{2j} (x)  \} =  \{ \Phi_{3} (y), \Phi_{2j} (x)  \} =  \{ \Phi_{4} (y), \Phi_{2j} (x)  \} =  \{ \Phi_{5i} (y), \Phi_{2j} (x)  \} \notag \\
&  \ \ \  \ \ \ \ \ \ \ \ \ \ \ \ \ \ \ \   =   \{ \Phi_{6} (y), \Phi_{2j} (x)  \} =  \{ \Phi^{(G)}_{Ni} (y), \Phi_{2j} (x)  \} =  \{ \Phi^{(G)}_{ei} (y), \Phi_{2j} (x)  \} =0 \ .
\end{align}
In order to classify the class of the constraints, it is more appropriate to choose 
\be
(\Phi_{1} , \Phi_{2j} , \Phi_{3} , \Phi_{4} , \tilde{\Phi}_{5j} , \Phi_{6})\label{constraintset}
\ee 
as a set of independent constraints. $\tilde{\Phi}_{5j} $ is defined in (\ref{Phi5tilde}) and given by a linear combination $\Phi_{5i}-\del_i N\Phi_1-\del_i e\Phi_3$.
As is clear from the above computation, $\Phi_{2j}$ and $\tilde{\Phi}_{5j}$ commute with all the constraints: 
\begin{align}
\{ \Phi_{1} (y), \Phi_{2j} (x)  \} &=   \{ \Phi_{2i} (y), \Phi_{2j} (x)  \} =  \{ \Phi_{3} (y), \Phi_{2j} (x)  \}=  \{ \Phi_{4} (y), \Phi_{2j} (x)  \} \nn\\
& =  \{ \tilde{\Phi}_{5i} (y), \Phi_{2j} (x)  \}   
=   \{ \Phi_{6} (y), \Phi_{2j} (x)  \}  =0 \ . \label{CA1}\\
 \{ \Phi_{1}(y) , \tilde{\Phi}_{5j}(x)   \} &=  \Phi_{1}(x) \partial_{y^{j}} \delta (y-x) \approx 0 \ , \\
\{ \Phi_{3}(y) , \tilde{\Phi}_{5j}(x)   \} &=  \Phi_{3}(x) \partial_{y^{j}} \delta (y-x) \approx 0\ , \\
\{ \Phi_{4} (y) , \tilde{\Phi}_{5j} (x)   \} &= \Phi_{4} (x) \partial_{y^{j}} \delta (y-x) \approx 0 \ , \\
\{ \tilde{\Phi}_{5i}(y) , \tilde{\Phi}_{5j}(x) \} &=\tilde{\Phi}_{5i}(x)\partial_{y^{j}}\delta (y-x) - \tilde{\Phi}_{5j}(y)\partial_{x^{i}}\delta (y-x) \approx 0 \ , \\
 \{ \Phi_{6} (y) , \tilde{\Phi}_{5j} (x)   \} &= \Phi_{6} (x) \partial_{y^{j}} \delta (y-x) \approx 0 \ . 
\label{CA6}
\end{align}  
It is easy to show that the set (\ref{constraintset}) is complete. Namely, the time flows of the secondary constraints do not give rise to any new constraints:
 \begin{align}
& \dot{\Phi}_{4}(x) = \int d^{3}y\{   \mathcal{H}^{e(0)}_{\text{nDBI}}(y)  ,\Phi_{4}(x) \} + \sum_{a = 1,3} \int d^{3}y \{   \Phi_{a}(y) , \Phi_{4} (x)  \} \lambda_{a} + \int d^{3}y \{  \Phi^{i}_{2}(y)   ,\Phi_{4} (x) \}\lambda_{2i}  \notag \\
&  \ \ \ \ \ \ \   \approx 0 \ , \label{ap1} \\
& \dot{\tilde{\Phi}}_{5j}(x) = \int d^{3}y\{   \mathcal{H}^{e(0)}_{\text{nDBI}}(y)  ,\tilde{\Phi}_{5j}(x) \} + \sum_{a = 1,3} \int d^{3}y \{   \Phi_{a}(y) , \tilde{\Phi}_{5j} (x)  \} \lambda_{a} + \int d^{3}y \{  \Phi^{i}_{2}(y)   , \tilde{\Phi}_{5j} (x) \}\lambda_{2i}  \notag \\
&  \ \ \ \ \ \ \   \approx 0 \ , \label{ap2} \\
& \dot{\Phi}_{6}(x) = \int d^{3}y\{   \mathcal{H}^{e(0)}_{\text{nDBI}}(y)  ,\Phi_{6}(x) \} + \sum_{a = 1,3} \int d^{3}y \{   \Phi_{a}(y) , \Phi_{6} (x)  \} \lambda_{a} + \int d^{3}y \{  \Phi^{i}_{2}(y)   ,\Phi_{6} (x) \}\lambda_{2i} \notag \\
&  \ \ \ \ \ \ \        \approx 0 \ . \label{ap3}
\end{align}
Since the Hamiltonian density takes the form
\begin{equation}
\mathcal{H}^{e(0)}_{\text{nDBI}} = -  \left( N \Phi_{4} + N^{j}  \Phi_{5j}  \right) - \frac{2}{\kappa} \sqrt{h} \left( e\Delta N - N \Delta e   \right) \ ,
\end{equation}
one can see that (\ref{ap1}) and (\ref{ap3}) determine $\lambda_{1}$ and $\lambda_{3}$, while (\ref{ap2}) is trivially satisfied. Hence, there are no additional constraints from these time flows. 
Having established the completeness of the set (\ref{constraintset}),  we conclude that $\Phi_{2j}$ and $\tilde{\Phi}_{5j}$ are first class, and the rest are second class.
Finally, let us end this appendix with a comment on the generator of the spatial diffeomorphism. 
The generator ${\cal G}(\xi^{i})$ of the spatial diffeomorphism acts on a phase-space variable $A$ as
\begin{equation}
\{ A(y) , {\cal G}(\xi^{i})  \} = \mathsterling_\xi A(y) \ .
\end{equation} 
Since $p_N, p^i_{\vec{N}}$, and $p_e$ are primary constraints, we only need to consider the reduced set of phase-space variables, $(h_{ij}, p^{ij}, N, \vec{N}, e)$. The spatial diffeomorphisms for this set are generated by $(\Phi_{5}, -\Phi_{5}, \Phi^{(G)}_{N}, \Phi^{(G)}_{\vec{N}},  \Phi^{(G)}_{e})$. They are indeed all generated by the first class constraints, as can be seen from (\ref{GN})--(\ref{Phi5tilde}).


\section{Computational details of perturbations}
\label{perturbation_background}

In this appendix, we show some details of the computations in the perturbative analysis of the scalar mode in Sections \ref{pathologies1} and \ref{conclusion}.

\subsection{Perturbation of the equations of motion}
\label{perturbation_eq_motion}
The linearised version of equations (\ref{EQmetric})--(\ref{EQevol}) and (\ref{EQscalar}) can be obtained in a fashion similar to \cite{Blas:2009yd}. To the approximation explained in Section \ref{pathologies1}, we find, in the gauge $N^i=0$,
\begin{align}
&\dot{\gamma}_{ij}-2\bar{N}\kappa_{ij}-2\bar{K}_{ij}n=0\ ,\label{PERTmetric}\\
&2\bar{K}_{ij}\kappa^{ij}-2\kappa\bar{K}-\nabla^i\nabla^j\gamma_{ij}+\Delta\gamma=\frac{G_N}{6\lambda}\Delta\alpha \ ,\label{PERTham}\\
&\nabla^j\kappa_{ij}-\nabla_i\kappa-\frac{3}{2}\bar{K}^{jk}\nabla_i\gamma_{jk}+\bar{K}^{jk}\nabla_k\gamma_{ij}+\frac{1}{2}\bar{K}_{ij}\nabla^j\gamma=\frac{G_N}{12\lambda}\left(\bar{K}_{ij}-\bar{h}_{ij}\bar{K}\right)\nabla^j\alpha\label{PERTmom}\ ,\\
&\dot{\kappa}^{ij}+\dot{\gamma}^{ij}\bar{K}-\bar{h}^{ij}\dot{\kappa}-\dot{\gamma}_{kl}\bar{h}^{ij}\bar{K}^{kl}-\nabla^i\nabla^jn+\bar{h}^{ij}\Delta n\nonumber\\
&\hspace{3cm}
+\frac{N}{2}\left(\nabla_k\nabla^j\gamma^{ik}+\nabla_k\nabla^i\gamma^{jk}
-\Delta\gamma^{ij}-\nabla^i\nabla^j\gamma\right)
=-\frac{G_N}{12\lambda}\bar{N}\nabla^i\nabla^j\alpha \ ,\label{PERTevol}
\end{align}
and 
\begin{align}
\hspace{-.0cm}
\Delta\dot{\alpha}=\Delta\del_t\left[\nabla^i\nabla^j\gamma_{ij}-\Delta\gamma+2\bar{K}_{ij}\kappa^{ij}-2\kappa\bar{K}-2\bar{N}^{-1}\left(\Delta n-\nabla_i\gamma^{ij}\nabla_j\bar{N}+\frac{1}{2}\nabla^i\gamma\nabla_i\bar{N}\right)\right]=0 \ .\label{PERTscalar}
\end{align}
The L.H.S. of (\ref{PERTmetric})--(\ref{PERTevol}) are the same as those in GR and agree with the $\lambda=\xi=1$ case of \cite{Blas:2009yd}. Eq.(\ref{PERTscalar}), however, is very different. 

We are interested in perturbations of the modes with wavelengths much shorter than the characteristic scale $L$ of the background, {\it i.e.}, $\omega,p\gg 1/L$, where the space-time is virtually flat.
First, (\ref{PERTscalar}) enforces $\alpha$ to the constant. Thus the R.H.S. of (\ref{PERTmetric})--(\ref{PERTevol}) are always negligible and thus we have in Fourier space
\begin{align}
&i\omega\gamma^{ij}+2(\bar{N}\kappa^{ij}+\bar{K}^{ij}n)=0\ ,\label{pert_metric}\\
&(p_ip_j-p^2\delta_{ij})\gamma^{ij}-2(\bar{K}\kappa-\bar{K}_{ij}\kappa^{ij})=0\ ,\label{pert_ham}\\
&p_j\kappa^{ij}-p^i\kappa-\frac{3}{2}\bar{K}_{jk}\gamma^{jk}p^i+\bar{K}_{jk}p^k\gamma^{ij}+\frac{1}{2}\bar{K}^{ij}p_j\gamma=0\ ,\label{pert_mom}\\
&i\omega(\delta^{ij}\kappa-\kappa^{ij})-(p^2\delta^{ij}-p^ip^j)n-i\omega(\bar{K}\gamma^{ij}-\delta^{ij}\bar{K}_{kl}\gamma^{kl})\nn\\
&\hspace{4.5cm}-\frac{N}{2}(p_kp^j\gamma^{ik}+p_kp^i\gamma^{jk}-p^2\gamma^{ij}-p^ip^j\gamma)=0\label{pert_evol}\ .
\end{align}
Plugging (\ref{pert_metric}) into (\ref{pert_evol}) and discarding sub-leading terms linear in  $\bar{K}_{ij}\sim{\cal O}(1/L)$, we simply obtain the fluctuation equation of GR in flat space-time:
\begin{equation}
\frac{\omega^2}{p^2}(\delta^{ij}\kappa-\kappa^{ij})+i\omega\left(\delta^{ij}-\frac{p^ip^j}{p^2}\right)n+\bar{N}^2\left(\kappa^{ij}+\frac{1}{p^2}\left(p^ip^j\kappa-p_kp^j\kappa^{ik}-p_kp^i\kappa^{jk}\right)\right)=0 \ .\label{pert_evol2}
\end{equation}
Contraction with $\delta_{ij}$ and $p_j$, respectively, yields
\begin{align}
i\omega\kappa-p^2n=&0\ ,\label{pert_evol2.1}\\
p^i\kappa-p_j\kappa^{ij}=&0\label{pert_evol2.2}\ .
\end{align} 
Using these relations, one can find that  (\ref{pert_evol2}) gives the massless dispersion relation
\begin{equation}
\omega^2=\bar{N}^2p^2\ .
\end{equation}
In this approximation, the rest of the equations, the Hamiltonian and momentum constraints  (\ref{pert_ham}) and (\ref{pert_mom}), are automatically satisfied. In GR, we can set $n=0$ by using the residual gauge symmetry. Thus we are left with a massless transverse traceless  tensor mode, {\it i.e.}, the usual graviton.

In $n$-DBI gravity, we do not have liberty to gauge away $n$. However,  
 we have (\ref{PERTscalar}) to take into account. Using (\ref{pert_metric}), it becomes
\begin{align}
&\kappa^{ij}
\left[i\omega\bar{N}\left(\bar{K}_{ij}-\bar{K}\delta_{ij}\right)+\bar{N}^2\left(p^2\delta_{ij}-p_ip_j+\delta_{ij}ip^k\partial_k\ln\bar{N}-2ip_i\partial_j\ln\bar{N}\right)\right]\nonumber\\
&\hspace{2cm}+n\left[-i\omega p^2+\bar{N}\bar{K}_{ij}\left(p^2\delta_{ij}-p_ip_j+ip^k\partial_k\ln\bar{N}\delta_{ij}-2ip_i\partial_j\ln\bar{N}\right)\right]=0\ .
\label{pert_scalar}
\end{align}
To leading order, this yields
\be
\bar{N}^2\kappa^{ij}\left(p^2\delta_{ij}-p_ip_j\right)-in\omega p^2=0\qquad\qquad
\stackrel{(\ref{pert_evol2.2})}{\Longrightarrow}\qquad\qquad n\omega p^2=0\ .
\ee
Since our approximation is valid only for $\omega, p\gg 1/L$, this implies that $n=0$ and we are again left with a graviton.
To the next order, however, this reads
\begin{align}
\hspace{-.5cm}
\kappa^{ij}\!\left[i\omega\bar{N}\left(\bar{K}_{ij}-\bar{K}\delta_{ij}\right)+\bar{N}^2\left(\delta_{ij}ip^k\partial_k\ln\bar{N}-2ip_i\partial_j\ln\bar{N}\right)\right]+
n\!\left[-i\omega p^2+\bar{N}\bar{K}_{ij}\left(p^2\delta_{ij}-p_ip_j\right)\right]=0\ .
\end{align}
Due to (\ref{pert_evol2.1}), the $\kappa^{ij}$ terms contribute. 
For $n$ to be non-vanishing, we must have the relation
\begin{equation}
\omega\sim \bar{N}p\sim  i\left(\bar{N}\bar{K}+\bar{N}\del\ln\bar{N}\right)\sim\frac{i}{L} \ .
\end{equation}
However, once again, this cannot be satisfied. Hence there is no scalar mode of the type $(n, \gamma)\sim \left(n(\omega, p), \gamma(\omega, p)\right)e^{i\omega t +ip\cdot x}$ with $|\omega|, |p|\gg 1/L$, including the one with imaginary $\omega$.

\subsection{Perturbation of the St\"uckelberg field}
\label{perturbation_stuckelberg}
To obtain the equation of motion for the St\"uckelberg field $\phi$, we vary the action (\ref{DBIgravitycovariant}) or (\ref{DBIgravitycovariante}) 
\begin{equation}
\delta S\sim e\delta\mathcal{K}\sim e\delta \left(D_\alpha\left(n^\alpha D_\beta n^\beta\right)\right)\sim\left(D_\alpha\left(n^\beta D_\beta e\right)-K D_\alpha e\right)\delta n^\alpha 
\ ,
\end{equation}
where $e=\left(1+\frac{G_N}{6\lambda}\left(\mbox{}^{(4)}\! R+\mathcal{K}\right)\right)^{-\half}$ and ${\cal K}=-2D_\alpha(n^\alpha D_\beta n^\beta)$. Using
\begin{equation}
\delta n^\alpha=\frac{\partial^\alpha\delta\phi+n^\alpha n^\beta\partial_\beta\delta\phi}{\sqrt{-X}}\ ,
\end{equation}
we find 
\begin{equation}
D^\alpha\left(\frac{(\partial_\alpha+n_\alpha n^\sigma \del_\sigma)(n^\beta \del_\beta e)-D_{\beta}n^{\beta}(\partial_\alpha+n_\alpha n^\sigma \del_\sigma)e}{\sqrt{-X}}\right)=0\ .\label{eq_scalar}
\end{equation}
It can be shown that this becomes (\ref{EQscalar}) when ${\bf n}=(-N,0,0,0)$, {\it i.e.}, $\phi=t$. Note that this equation is invariant under $\phi\to f(\phi)$, as it should be, owing to the fact that $n^{\alpha}(\partial_\alpha+n_\alpha n^\sigma \del_\sigma)=0$. This also implies that the quantity in the parenthesis is proportional to the space-like vector $n^{\beta}D_{\beta}n_{\alpha}$ tangential to the hypersurface.

Now we consider perturbations $\phi=\bar{\phi}+\varphi$, 
expanding (\ref{eq_scalar}) to linear order in $\varphi$. We work in unitary gauge, $\bar{\phi}=t$ and $N^i=0$ and use
\begin{equation}
\delta n^{\alpha}=(0,\bar{N}\partial^i\varphi)\ ,\qquad \delta\left(
\sqrt{-X}\right)^{-1/2}=-\bar{N}\dot{\varphi}\ , \qquad \delta K=\bar{N}^{-1}\nabla_i(\bar{N}^2\del^i\varphi)\ ,
\end{equation}
as well as
\begin{eqnarray}
\delta e=-\frac{G_N}{12\lambda}\bar{e}^3\delta\mathcal{K}\ ,\qquad
-\frac{1}{2}\delta\mathcal{K}=(2\bar{K}-\bar{N}^{-1}\del_t)(\bar{N}^{-1}\nabla_i(\bar{N}^2\del^i\varphi))+\bar{N}\del^i\varphi\del_i \bar{K}\ .
\label{evarphirelation}
\end{eqnarray}
The terms coming from perturbing $n^\alpha$ and $1/\sqrt{-X}$ in (\ref{eq_scalar}) contain at most 3 $\varphi$-derivatives, while those from perturbing $e$ contain 6 and 5 $\varphi$-derivatives. Thus we only need to consider the latter. It also suffices to keep 3 and 2 $\delta e$-derivatives. Then we find
\begin{equation}
\Delta\delta\dot{e}+\bar{K}\bar{N}\Delta\delta e=0\ .
\end{equation}
In terms of $\varphi$ the leading terms (with 6 and 5 derivatives) yield (\ref{phi3}).

\subsection{Perturbation of the St\"uckelberg field - 2nd order}
\label{perturbation_stuckelberg2}
To find the cubic action of the fluctuation $\varphi$ in flat space-time (with $q=1$), we expand the action as
\be
S=-{1\over 16\pi G_N}\int d^4x
\left[{\cal K}-{1\over 4}\left({G_N\over 6\lambda}\right){\cal K}^2-{5\over 8}\left({G_N\over 6\lambda}\right)^2{\cal K}^3+\cdots\right]\ .\label{Scubicchi}
\ee
The first term is a surface term and does not contribute to the equation of motion. In unitary gauge,  the time-like vector takes the form
\begin{align}
n_0=-{1+\dot{\varphi}\over\sqrt{\left(1+\dot{\varphi}\right)^2-(\del_i\varphi)^2}}\ ,\qquad\qquad
n_i=-{\del_i\varphi\over\sqrt{\left(1+\dot{\varphi}\right)^2-(\del_i\varphi)^2}}\ .
\label{timelikevector}
\end{align}
This can be expanded as
\begin{align}
&-n_0=1+\half(\del_i\varphi)^2
-\dot{\varphi}(\del_i\varphi)^2+{\cal O}(\varphi^4)\ ,\\
&-n_i=\del_i\varphi-\dot{\varphi}\del_i\varphi
+\dot{\varphi}^2\del_i\varphi+\half(\del_i\varphi)^3+{\cal O}(\varphi^4)\ .
\end{align}
In the flat background we have
\begin{align}
{\cal K}
=-2\left[\del_0\left(n^0\del_0n^0\right)+\del_i\left(n^i\del_jn^j\right)
+\del_0\left(n^0\del_in^i\right)+\del_i\left(n^i\del_0n^0\right)\right]\ ,
\end{align}
and we find to quadratic order
\begin{align}
{\cal K}=2\left[\Delta\dot{\varphi}+\del_0(\dot{\varphi}\Delta\varphi)
-\del_i(\del_i\varphi\Delta\varphi)\right]\ .
\end{align}
Hence the cubic scalar field action is given by
\begin{align}
S_3={1\over 192\pi \lambda}\int d^4x
\left[\half(\Delta\dot{\varphi})^2-\left(\dot{\varphi}\Delta\ddot{\varphi}
-\del_i\varphi\Delta\del_i\dot{\varphi}\right)\Delta\varphi
+{5G_N\over 2\lambda}{1\over 3!}(\Delta\dot{\varphi})^3\right]\ .
\label{cubic}
\end{align}

\subsection{A nonlinear analysis of the St\"uckelberg field}
\label{nonlinear_stuckelberg}

The scalar mode $\varphi$ obeys the equation of motion  (\ref{eq_scalar}). To determine whether the scalar mode leads to an instability or not, we need to study  (\ref{eq_scalar}) beyond linear order approximation.
As we discussed, to linear order, the equation of motion is simply
\be
\Delta^2\ddot{\varphi}=0\ ,
\ee
and the solution is 
\be
\varphi(t,x)=\epsilon\left[\varphi_0(x)+\varphi_1(x) t\right]\ ,
\ee
where $\varphi_0(x)$ and $\varphi_1(x)$ are arbitrary functions of space. We have included the factor of $\epsilon\ll 1$ for later convenience. The fully nonlinear solution can in principle be found systematically order by order in $\epsilon$ expansions:
\be
\varphi(t,x)=\epsilon\left[\varphi_0(x)+\varphi_1(x) t\right] + \sum_{n=2}^{\infty}
\epsilon^n\varphi_n(t,x)\ .\label{series_app}
\ee
The higher order fluctuations $\varphi_n(t,x)$'s are determined in terms of the initial data $\left(\varphi_0(x), \varphi_1(x)\right)$ and of order $n$ in powers of (spatial derivatives of) $\varphi_0(x)$ and $\varphi_1(x)$ and polynomial in time $t$.
Using (\ref{cubic}), for example,  the next-to-leading order fluctuation can be found as
\be
\varphi_2(t,x)=\half\varphi_2^{(2)}(x)t^2+{1\over 3!}\varphi_2^{(3)}(x)t^3\ ,
\label{nexttoleadingsol}
\ee
where
\begin{align}
\Delta^2\varphi_2^{(2)}=&\Delta\left[2\Delta\varphi_0\Delta\varphi_1+2\del_i\varphi_0\Delta\del_i\varphi_1
+\Delta\del_i\varphi_0\del_i\varphi_1\right]
-\left(\Delta\varphi_0\Delta^2\varphi_1+\Delta\del_i\varphi_0\Delta\del_i\varphi_1\right)\ ,
\nn\\
\Delta^2\varphi_2^{(3)}=&\Delta\left[2(\Delta\varphi_1)^2+2(\del_i\varphi_1)^2
+\del_i\varphi_1\Delta\del_i\varphi_1\right]-\left(\Delta^2\varphi_1\Delta\varphi_1
+\del_i\varphi_1\Delta\del_i\varphi_1\right)\ .\label{phi2coefficients}
\end{align}
Clearly, the late time behaviour of the scalar mode requires the knowledge of all order fluctuations. Thus, to see whether the scalar mode yields an instability or not, we need to resum the infinite series (\ref{series_app}). However, this seems to be out of our reach and we will instead resort to an alternative analysis working in the Einstein frame.

\section{Linear perturbations in the Einstein frame}
\label{einstein_frame}

We consider linear perturbations of the scalar fields around flat space-time in the Einstein frame (\ref{Einstein}). The scalar field Lagrangian density reads
\begin{align}
4\pi G_N\mathcal{L}^E_{\rm scalar} = & 2\dot{\psi}^2+4\dot{\psi}(\dot{E}+\Delta B)-(4n-2\psi)\Delta\psi\nn\\
&+4\dot{\chi}\left(\dot{E}+\Delta B+2\dot{\psi}\right)
+6\dot{\chi}^2-6\chi\Delta\chi-\frac{24\lambda}{G_N}\chi^2\ .
\end{align}
This is the Einstein frame counterpart of (\ref{scalarL}). The equations of motion are given by
\begin{align}
\ddot{\psi} &= -\ddot{\chi}\ ,\\
\Delta\dot{\psi} &= -\Delta\dot{\chi}\ ,\\
\ddot{E}+\Delta\dot{B}+\Delta n &= -\ddot{\chi}\ ,\\
\Delta\psi &= 0\ ,
\end{align}
which clearly reduce to those of GR for constant $\chi$, plus
\begin{equation}
\ddot{\chi}+\Delta\chi+\frac{1}{3}(\ddot{E}+\Delta\dot{B}+2\ddot{\psi})+\frac{4\lambda}{G_N}\chi=0\ .
\label{chi_ln}
\end{equation}
This is the linearisation of (\ref{psi_eqn}). In the $E = 0$ gauge the general solution can be found as
\begin{align}
B= &  B_0(x)+B_1(x)t\ ,\\
n= & -B_1(x)\ ,\\
\psi= & 0\ ,\\
\chi = & \chi_0(x)\ ,
\end{align}
with $\chi_0(x)$ related to $B_1(x)$ by
\begin{align}
\Delta B_1=-3\Delta\chi_0-\frac{12\lambda}{G_N}\chi_0\ .
\end{align} 
Here we have again imposed the boundary condition that all the fields fall off at spatial infinity. 
Note that $\chi_0(x)$ is essentially $B_1(x)$ which is the degree of freedom responsible for the linear time growth. This suggests the identification $\chi(t,x) \sim \dot{T}(t,x)$, that is, the auxiliary field $\chi$ can be regarded as the time derivative of the scalar mode.


\end{document}